\documentclass[conference]{IEEEtran}
\IEEEoverridecommandlockouts

\usepackage[dvipsnames]{xcolor}
\usepackage{color}

\definecolor{Thesisgreen}{HTML}{3E0751}
\definecolor{Thesispink}{HTML}{5DB37F}

\usepackage[T1]{fontenc}
\usepackage[normalem]{ulem}
\usepackage{pgfplots}
\pgfplotsset{compat = newest}
\usepackage{adjustbox}
\usepackage[utf8]{inputenc}
\usepackage{physics}
\usepackage{siunitx}
\usepackage{cite}
\usepackage{textcomp}
\usepackage{bm} 
\usepackage{bbm} 
\usepackage{braket}
\usepackage{dsfont}
\usepackage{amsmath}
\usepackage{amsthm}
\usepackage{cite}
\usepackage{textcomp}
\usepackage{amssymb}
\usepackage{thmtools} 
\usepackage{mathtools}
\usepackage[switch]{lineno} 
\usepackage[ruled,linesnumbered]{algorithm2e}
\SetArgSty{textnormal}
\usepackage{algpseudocode}
\usepackage{url}

\usepackage{subfiles}
\usepackage{tikz}
\usepackage{stmaryrd}
\usepackage{thm-restate}

\makeatletter
\newcommand{\gettikzxy}[3]{%
  \tikz@scan@one@point\pgfutil@firstofone#1\relax
  \edef#2{\the\pgf@x}%
  \edef#3{\the\pgf@y}%
}
\makeatother
\usepackage{graphics}
\usepackage{graphicx}
\usepackage{float}
\usepackage[colorlinks]{hyperref}
\hypersetup{
    colorlinks  = true,
    citecolor   = RoyalPurple,
    linkcolor   = teal,
    urlcolor    = teal
}

\apptocmd{\thebibliography}{\small}{}{}

\declaretheorem[style=plain]{theorem}

\declaretheorem[style=plain,sibling=theorem]{lemma}

\declaretheorem[style=definition,sibling=theorem]{definition}

        \DeclareMathOperator{\sign}{sign}
        \DeclareMathOperator{\QSVT}{QSVT}
        \DeclareMathOperator{\QSP}{QSP}
        \DeclareMathOperator{\Amplify}{Amplify}
        \DeclareMathOperator{\rect}{rect}
        
        \newcommand*{\N}{\mathbb{N}}

        \newcommand*{\R}{\mathbb{R}}

        \newcommand{\cost}{\ensuremath{\mathcal{Q}}} 
        \newcommand{\qlsF}{\mathrm{QLS_F}}
        \newcommand{\qlsC}{\mathrm{QLS_C}}
        \newcommand{\qlsQ}{\mathrm{QLS_{QSVT}}}
        \newcommand{\lcu}{\mathrm{LCU}}
        \newcommand{\qpe}{\mathrm{QPE}} 
        \newcommand{\dom}[1]{\mathcal{D}_{#1}} 
        \newcommand{\nqaa}{n_{\mathrm{QAA}}}

\newcommand{\bor}{\ensuremath{\mathcal{P}_b}}

\usepackage{tcolorbox}  
\usepackage{ragged2e}
\usepackage{enumitem}

\def\BibTeX{{\rm B\kern-.05em{\sc i\kern-.025em b}\kern-.08em
    T\kern-.1667em\lower.7ex\hbox{E}\kern-.125emX}}
\selectfont

\begin{document}
\title{Beyond asymptotic scaling: \\Comparing functional quantum linear solvers}

\author{
\IEEEauthorblockN{
    Andreea-Iulia Lefterovici\IEEEauthorrefmark{1}${}^{,}$\IEEEauthorrefmark{2}, 
    Michael Perk\IEEEauthorrefmark{3}, 
    Debora Ramacciotti\IEEEauthorrefmark{1}
    Antonio F.\ Rotundo\IEEEauthorrefmark{4},\\ 
    S. E. Skelton\IEEEauthorrefmark{1}${}^{,}$\IEEEauthorrefmark{2}, 
    Martin Steinbach\IEEEauthorrefmark{1}
}
\IEEEauthorblockA{\IEEEauthorrefmark{1}Institut f\"ur Theoretische Physik, Leibniz Universit\"at Hannover, Hannover, Germany}
\IEEEauthorblockA{\IEEEauthorrefmark{2}\{andreea.lefterovici, shawn.skelton\}@itp.uni-hannover.de}
\IEEEauthorblockA{\IEEEauthorrefmark{3}Department of Computer Science, Technische Universit\"at Braunschweig, Braunschweig, Germany}
\IEEEauthorblockA{\IEEEauthorrefmark{4}Dipartimento di Fisica, Universit\`{a} di Pavia, Pavia, Italy}
}

\maketitle
\thispagestyle{plain}
\pagestyle{plain}

\begin{abstract}
Solving systems of linear equations is a key subroutine in many quantum algorithms. In the last 15 years, many quantum linear solvers (QLS) have been developed, competing to achieve the best asymptotic worst-case complexity.
Most QLS assume fault-tolerant quantum computers, so they cannot yet be benchmarked on real hardware.
Because an algorithm with better asymptotic scaling can underperform on instances of practical interest, the question of which of these algorithms is the most promising remains open.
In this work, we implement a method to partially address this question. We consider four well-known QLS algorithms which directly implement an approximate matrix inversion function: the Harrow-Hassidim-Lloyd algorithm, two algorithms utilizing a linear combination of unitaries, and one utilizing the quantum singular value transformation (QSVT). These methods, known as functional QLS, share nearly identical assumptions about the problem setup and oracle access. 
Their computational cost is dominated by query calls to a matrix oracle encoding the problem one wants to solve. 
We provide formulas to count the number of queries needed to solve specific problem instances; these can be used to benchmark the algorithms on real-life instances without access to quantum hardware.
We select three data sets: random generated instances that obey the assumptions of functional QLS, linear systems from simplex iterations on MIPLIB, and Poisson equations. 
Our methods can be easily extended to other data sets and provide a
high-level guide to evaluate the performance of a QLS algorithm.
In particular, our work shows that HHL underperforms in comparison to the other methods across all data sets, often by orders of magnitude, while the QSVT-based method shows the best performance.
\end{abstract}



\section{\label{section:Introduction}Introduction}
Solving systems of linear equations has been a core interest of quantum algorithms research since the early 2000s.
Many proposed applications of quantum technology are reduced to some version of a quantum linear solver (QLS) algorithm~\cite{Liu2021RigorousRobustQuantumSpeedUpSupervisedMachineLearning, Schuld2021SupervisedQuantumMachineLearningKernelModels}, especially within the field of quantum machine learning \cite{schuld2016prediction, wiebe2012quantum, rebentrost2014quantum, Wang2017QuantumAlgorithmForLinearRegression, shao2018reconsiderhhlalgorithmrelated, lloyd2013quantumalgorithmssupervisedunsupervised}.
Harrow, Hassidim, and Lloyd proposed the first QLS, the HHL algorithm \cite{Harrow2009QuantumAlgorithmLinearSystemsEquations}, showing an exponential time improvement\footnote{This should be taken with a grain of salt, as the solution is encoded in a quantum state, hence we do not obtain as much information as in the classical case.} in the problem size over classical asymptotically optimal algorithms such as the conjugate gradient method \cite{Shewchuk1994IntroductionToConjucateeGradiientMethod}.
HHL has an overall asymptotic complexity of $\mathcal{O}(\log(N) d^2 \kappa^2 / \varepsilon)$, where $N$ is the number of variables, $d$ is the maximum sparsity per row or column, $\kappa$ is the condition number and $\varepsilon$ is the desired solution accuracy.

Subsequent improvements to HHL have refined the asymptotic time complexity to nearly
$\mathcal{O}(\kappa d \log(\kappa/\varepsilon))$ \cite{Ambainis2012VariableTimeAmplitudeAmplification, Childs2017QuantumAlgorithmSistemyLinearEquations, Gilyen2019QuantumSingularValueTransformation, Subasi2019QuantumAlgorithmSystemsLinearEquationsAdiabaticQuantumComputing,Dong2022QuantumLinearSysteSolverAdiabationQuantumComputingQAOA, Costa2022OptimalScalingQuantumLinearSystemsSolverDiscreteAdiabaticTheorem}.
Here, we consider and compare four of these algorithms: HHL, two algorithms based on linear combination of unitaries \cite{Childs2017QuantumAlgorithmSistemyLinearEquations} which we denote QLS-Fourier and QLS-Chebyshev, and one based on QSVT \cite{Gilyen2019QuantumSingularValueTransformation}, which we denote QLS-QSVT. 
These algorithms belong to the class of functional quantum linear solvers because they implement a matrix function of $A$ which approximates the matrix inverse $A^{-1}$.
They use essentially the same resources: sparse matrix oracles, Hamiltonian simulation, and variations of
quantum amplitude amplification (QAA). 
We focus on these algorithms because functional QLS, especially the HHL algorithm, are among the most popular and well-understood methods to date.

It is important to compare algorithms with different asymptotic costs, because an algorithm with worse asymptotic scaling can still outperform one with a better theoretical scaling due to, for example, stability issues.
A well-known example is the ellipsoid method~\cite{Khachiyan1979APolynomialAlgorithmInLinearProgramming} versus the simplex method \cite{danzig1948linear} in classical optimization. 
This perspective justifies considering HHL; while its asymptotic complexity suggests that it is less efficient, its practical performance could have been different.

An alternative approach is to use methods closely related to adiabatic quantum computing to create a QLS \cite{Subasi2019QuantumAlgorithmSystemsLinearEquationsAdiabaticQuantumComputing}. 
Combined with an eigenvalue filtering routine from \cite{Dong2022QuantumLinearSysteSolverAdiabationQuantumComputingQAOA}, this approach has led to a quantum algorithm with optimal scaling in the accuracy and condition number \cite{Costa2022OptimalScalingQuantumLinearSystemsSolverDiscreteAdiabaticTheorem}.
Nevertheless, adiabatic QLS are structurally distinct from functional QLS, and therefore more difficult to compare with the same assumptions.

As quantum algorithms advance toward intermediate-term applications, it is becoming increasingly important to understand which quantum algorithms might be most successful in \textit{relevant parameter regimes and practical use cases}.
While the asymptotic performance of QLS methods is well understood, a practical comparison requires analyzing their exact query costs for specific condition number, sparsity levels, and accuracy requirements.

Our goal is to provide a comparative analysis of different functional QLS, as understanding the performance and their practical applicability is important for guiding future software developments, even before fault-tolerant quantum hardware is available.
We explore a relatively new approach to resource estimation that was originally introduced in \cite{Cade2022QuantifyingGroverSpeedupsBeyondAsymptoticAnalisys}.
We are using a version of this hybrid benchmarking method by counting the exact number of query calls to the same oracles (detailed in \autoref{section:Preliminaries}). 
Since we are interested in comparing quantum algorithms, an end-to-end cost analysis is not required. 
We assume the same sparse access oracles for all algorithms and compare them by counting the number of required queries. 
In particular, we do not consider quantum error correction, because this would give the same overhead to all algorithms.
This allows us to clarify which quantum algorithm succeeds best for the problems we consider, and to explore applications for fault-tolerant algorithms without requiring access to powerful simulators or quantum devices.
Further variations of this hybrid benchmarking method were used in \cite{Ammann2023RealisticRuntimeAnalysisQuantumSimplex, Wagner2024QuantumSubroutinesInBranchPrinceAndCutForVechicleRouting, Brehm2024AssessingFTQAdvantageForKSATWithStructure, henze2025solvingquadraticbinaryoptimization, ostermann2025benchmarkingquantumclassicalsdp}. 

Our analysis shows that HHL consistently underperforms in comparison to the other methods over all instance sets, often by several orders of magnitude. 
This large difference shows the limitations of HHL as a practical QLS, suggesting that other approaches may work better.
Among the tested methods, QLS-QSVT consistently performs better than the others across all data sets, making it the overall winner.
Assessing the performance of QLS before fault-tolerant hardware is ready shows the usefulness of the hybrid benchmarking methods developed in this work.


\section{\label{section:Preliminaries}Preliminaries}
\subsection{Quantum Linear Solvers}\label{subsection:PreliminariesQuantumLinearSolvers}
We consider linear systems of equations 
\begin{equation*}
    Ax=b \, ,
\end{equation*}
where we assume for simplicity that $A \in \mathbb{C}^{N\times N}$ is a $N \times N$ invertible Hermitian matrix, and $b \in \mathbb{C}^N$ a complex vector. Assume that $A$ is $d$-sparse, that is, $A$ has at most $d$ nonzero entries in any row or column. 
For Hermitian matrices, the condition number $\kappa$ is the ratio between the largest and the smallest eigenvalue of $A$ as $\kappa(A) = \frac{\abs{\lambda_{\text{max}}(A)}}{\abs{\lambda_{\text{min}}(A)}}$.
Finally, we say that a sequence of square matrices $(A_j)$ with sizes $(N_j) \subset \N$ is \emph{well-conditioned} if $\kappa(A_j) = \mathcal{O}(\text{poly}(\log N_j))$.

If not specified otherwise, the norm $\norm{\cdot}$ for vectors and matrices denotes the 2-norm.
We assume that the matrix $A$ and the vector $b$ have unit norm, which can always be enforced by classical pre-processing.  
Moreover, we assume that we have access to an oracle $\bor$ that prepares the quantum state $\ket{b}$,  
\begin{equation*}
\bor\ket{0}=\ket{b}=\sum_i b_i|i\rangle\,.
\end{equation*}
The goal of a quantum linear solver is to output a state $\ket{\tilde{x}}$ such that $\norm{\ket{\tilde{x}}-\ket{x}} \leq \varepsilon$, where
\begin{equation*}
    \ket{x} \coloneqq \frac{\sum_i x_i|i\rangle}{\| \sum_i x_i|i\rangle \|} \,
\end{equation*}
is a quantum state proportional to the solution $x=A^{-1}b$.

To prepare $A$, we assume that all the functional oracles discussed here rely on the sparse access input model (SAIM) \cite{Berry2015HamiltonianSimulationNearlyOptimalDependenceOnAllParameters, Childs2017QuantumAlgorithmSistemyLinearEquations}.
Assume the $j, k$-th element of $A$, $A_{jk}$, is given as a bit string representation. 
Then we define $\mathcal{P}_A$ consisting of two suboracles

\begin{align}
    &\mathcal{O}_F: \ket{j, l}\mapsto \ket{j, f(j, l)} ,
    \label{equation:SparseOracles}\\
    &\mathcal{O}_A: \ket{j, k, z}\mapsto\ket{j, k, z\oplus A_{jk}} ,
\end{align}
where $f$ is a function returning the column index of the $l^{th}$ non-zero element of row $j$ and the operation $\oplus$ indicates bit-wise addition modulo $2$, $z$ is the value of the register storing the bit string representation of $A_{jk}$. 
Note that QLS-Fourier and QLS-QSVT can also be applied \cite{Gilyen2019QuantumSingularValueTransformation, Childs2017QuantumAlgorithmSistemyLinearEquations} when some of the above conditions are relaxed, but for the purposes of joint comparison we require all of the assumptions to be enforced.

\subsection{Hybrid Benchmarking}
Asymptotic scaling is the standard measure for comparing the performance of quantum algorithms. 
However, in the case of quantum algorithms with similar asymptotic scaling, prefactors become relevant in predicting which algorithm might be preferable for practical instances with a given finite size. 
For systems of linear equations, the dependence on parameters such as accuracy, condition number, and sparsity makes assessing their performance outside of the asymptotic regime challenging. 
In this work, we address the following question:
Which quantum linear solver has the smallest number of query calls for several relevant real-world problems?

In our hybrid benchmarking technique, we derive estimates of the number of calls to the oracles $\mathcal{O}_F$ and $\mathcal{O}_A$ made by HHL, QLS-Fourier, QLS-Chebyshev and QLS-QSVT (see \autoref{section:Method}).
This is reasonable because the number of query calls to the oracles $\mathcal{O}_F$ and $\mathcal{O}_A$ dominates the cost\footnote{This is a common assumption in quantum algorithms literature. For QSVT, the number of 1- and 2-qubit gates is linear in the number of calls to the oracle, which is presumably far more expensive than other parts of the method\cite{Gilyen2019QuantumSingularValueTransformation}. Similar justifications for focusing on these oracle costs exist in \cite{Harrow2009QuantumAlgorithmLinearSystemsEquations, Childs2017QuantumAlgorithmSistemyLinearEquations}}.
While the costs of preparing $\ket{b}$ are non-trivial, they are expected to be comparable for every algorithm we consider, so we do not need to account for them.
We also neglect the quantum error correction overhead.

The linear systems we consider arise from randomly generated systems that fit the assumptions detailed in \autoref{subsection:PreliminariesQuantumLinearSolvers}, simplex iterations on MIPLIB \cite{miplib2021}, and Poisson equations with Dirichlet boundary conditions.
For all instances, we record on a classical computer the relevant problem parameters $\norm{x}$, $\norm{A}_{\text{max}}$, $\kappa$ and $d$. 
With these parameters, we calculate the number of oracle calls needed to solve each of these particular instances.


\section{\label{section:Method}Methods}
Functional quantum linear solvers implement the functional calculus $\frac{1}{x} \mapsto A^{-1}$ of a Hermitian operator $A$ by approximating $x \mapsto 1/x$.
These approximation functions are sums either over Hamiltonian simulation steps $e^{iAt}$ (HHL, QLS-Fourier), or Chebyshev polynomials of the first kind prepared with block encodings (QLS-Chebyshev, QLS-QSVT). 
However, the Hamiltonian simulation algorithm we use relies on the same quantum walk oracles as the block encoding for QLS-Chebyshev. 
We discuss this in detail in \autoref{sec:block_encodings}. 
All require some form of QAA to succeed with high probability: in our case always Alg. \ref{alg:qaa}.

We denote by $\cost[\ldots]$ the number of queries made by each algorithm to the oracles $\mathcal{O}_F$ and $\mathcal{O}_A$.

\begin{restatable}[Hamiltonian simulation by qubitization]{lemmma}{LemmaCostQubitization}\label{lemma:Cost_of_Qubitization}
    Given access to the oracles in \eqref{equation:SparseOracles} specifying a $d$-sparse Hamiltonian $A$, the algorithm of \cite{Low2019Qubitization} simulates $e^{-iAt}$ for $t\in\mathbb{R}_+$ and precision $\varepsilon\in(0, 1)$ using
    \begin{equation}\label{eq:ham_sim}
        \cost[e^{-iAt}] = 48\Tilde{r}\left(d||A||_{\text{max}}t, \varepsilon\right) = 48\Tilde{r}\left(\Tilde{t}, \varepsilon\right)
    \end{equation}    
queries to the oracles $\mathcal{O}_F$ and $\mathcal{O}_A$, where
\begin{equation*}
\begin{aligned}
        \Tilde{r}(\Tilde{t}, \varepsilon)=\begin{cases}
            \lceil e \Tilde{t}\rceil
            \quad  &\Tilde{t}\geq \frac{\ln(1/\varepsilon)}{e} \, ,\\
             \Bigl\lceil \frac{4\log(1/\varepsilon)}{\log\left(e+\frac{1}{\Tilde{t}}\log(1/\varepsilon)\right)} \Bigr\rceil
             \quad &\Tilde{t}< \frac{\ln(1/\varepsilon)}{e} \, .
        \end{cases}
\end{aligned}
\end{equation*}
Here, $\Tilde{t} = d||A||_{\text{max}}t$ is a rescaled version of $t$.
\end{restatable}

\begin{restatable}[QAA overhead]{lemmma}{LemmaCostQAA}\label{lemma:CostQAA}
Consider an algorithm $\mathcal{A}$ that is successful with probability $p=\sin^2{\theta}$, and let $p_0\le p$ be a known lower bound on this probability.
QAA boosts the probability of success to $\mathcal{O}(1)$ by applying algorithm $\mathcal{A}$ an expected number of times equal to
\begin{equation}\label{n_q}
 \nqaa=\sum_{k=1}^{\infty}\Bigl[m_k \prod_{l=1}^{k-1}\left(1-\left\langle p_{j_l}\right\rangle\right)\Bigr] = \mathcal{O} \left( \frac{1}{\sqrt{p}} \right) ,  
\end{equation}
where 
\begin{equation*}
\begin{split}
\left\langle p_{j_l}\right\rangle &=\frac{1}{2}-\frac{\sin(4(m_l + 1)\theta)}{4(m_l + 1) \sin (2 \theta)} \, ,\\
m_k &=  \lfloor\min(c^k, \sqrt{1/p_0})\rfloor \, ,
\end{split}
\end{equation*}
for a constant $1<c<2$.\footnote{Throughout the paper we pick $c=6/5$.
}
\label{lemma:CostQSearch}
\end{restatable}
We assume that when running QAA only the lower bound $p_0$ is known. In \textit{hybrid benchmarking}, we classically compute $\left|\left|x\right|\right|$ to calculate the exact probability $p$ and obtain the query calls estimate.

As $m_k$ is used in the quantum algorithm, it can only depend on $p_0$, while to compute $\langle p_{j_l}\rangle$, which is only required for the benchmark, we use the exact probability $p$ (through $\theta$).

\begin{restatable}[Query complexity of HHL]{lemmma}{LemmaCostHHL}
Given access to the oracles in \eqref{equation:SparseOracles} specifying a $d$-sparse Hamiltonian $A$, the algorithm of \cite{Harrow2009QuantumAlgorithmLinearSystemsEquations} prepares the state $A^{-1}\ket{b}$ up to normalization and with an accuracy of $\varepsilon$ using
\begin{equation*}
    \cost[\text{HHL}] = 2\nqaa\cdot\cost[e^{-iAt}] \, .
\end{equation*}  
queries to the oracles $\mathcal{O}_F$ and $\mathcal{O}_A$.
The QAA overhead $\nqaa$ is computed using Eq.\ \eqref{n_q}, using the following estimates for the probability of success $p$ and its lower bound $p_0$
\begin{equation*}
    p_0= \frac{1}{4\kappa^2} \, , \quad 
    p = \frac{\norm{x}^2}{4\kappa^2} \, ,
\end{equation*}
which are valid as long as $\varepsilon\ll 1/\kappa$.
Here, $x=A^{-1}b$ is the solution of the linear system of equations.
The Hamiltonian simulation is run for a time 
\begin{equation*}
    t = \sqrt{c_F} \kappa/\varepsilon \, ,
\end{equation*} with 
\begin{equation*}
    c_F=2^7+22\pi^2+(64+14\pi^2)^2/\pi^2  \, .
\end{equation*}
We have used Hamiltonian simulation by qubitization whose query complexity is given by Eq.\ \eqref{eq:ham_sim}.
\label{lemma:CostHHL}
\end{restatable}

\begin{restatable}[Query complexity of QLS-Fourier]{lemmma}{LemmaCostQLSF}
Given access to the oracles in \eqref{equation:SparseOracles} specifying a $d$-sparse Hamiltonian $A$, the algorithm of \cite{Childs2017QuantumAlgorithmSistemyLinearEquations} prepares the state $A^{-1}\ket{b}$ up to normalization and with an accuracy of $\varepsilon$ using
\begin{equation*}
\cost[\qlsF] = \nqaa\cdot\cost[e^{-iAt}] \, .
\end{equation*}
queries to the oracles $\mathcal{O}_F$ and $\mathcal{O}_A$.
The QAA overhead $\nqaa$ is computed using Eq.\ \eqref{n_q}, using the following expressions for the probability of success $p$ and its lower bound $p_0$
\begin{equation*}
    p_0= \frac{1}{\alpha^2} \, , \quad 
    p = \frac{\norm{x}^2}{\alpha^2} \, .
\end{equation*}
Here, $x=A^{-1}b$ is the solution of the linear system of equations, and 
\begin{equation*}
    \alpha=4\sqrt{\pi}\frac{\kappa}{\kappa+1}\sum_{l=1}^{L}l\Delta_z e^{-(l\Delta_z)^2/2}\,,
\end{equation*} 
with 
\begin{equation*}
\begin{split}
    \Delta_z =\frac{2\pi}{\kappa+1}\Bigl[\log\Bigl(1+\frac{8\kappa}{\varepsilon}\Bigr)\Bigr]^{-1/2}\,,
    L=\Bigl\lfloor\frac{\kappa+1}{\pi}\log\Bigl(1+\frac{8\kappa}{\varepsilon}\Bigr)\Bigr\rfloor\,.
\end{split}
\end{equation*}
The Hamiltonian simulation algorithm is run for a time
\begin{equation*}
    t=2\sqrt{2}\kappa \log\Bigl(1+\frac{8\kappa}{\varepsilon}\Bigr)\,.
\end{equation*}
We have used Hamiltonian simulation by qubitization whose query complexity is given by Eq.\ \eqref{eq:ham_sim}.
\label{lemma:CostQLS-Fourier}
\end{restatable}

\begin{figure*}[ht!]
    \centering
    \includegraphics[scale=0.52]{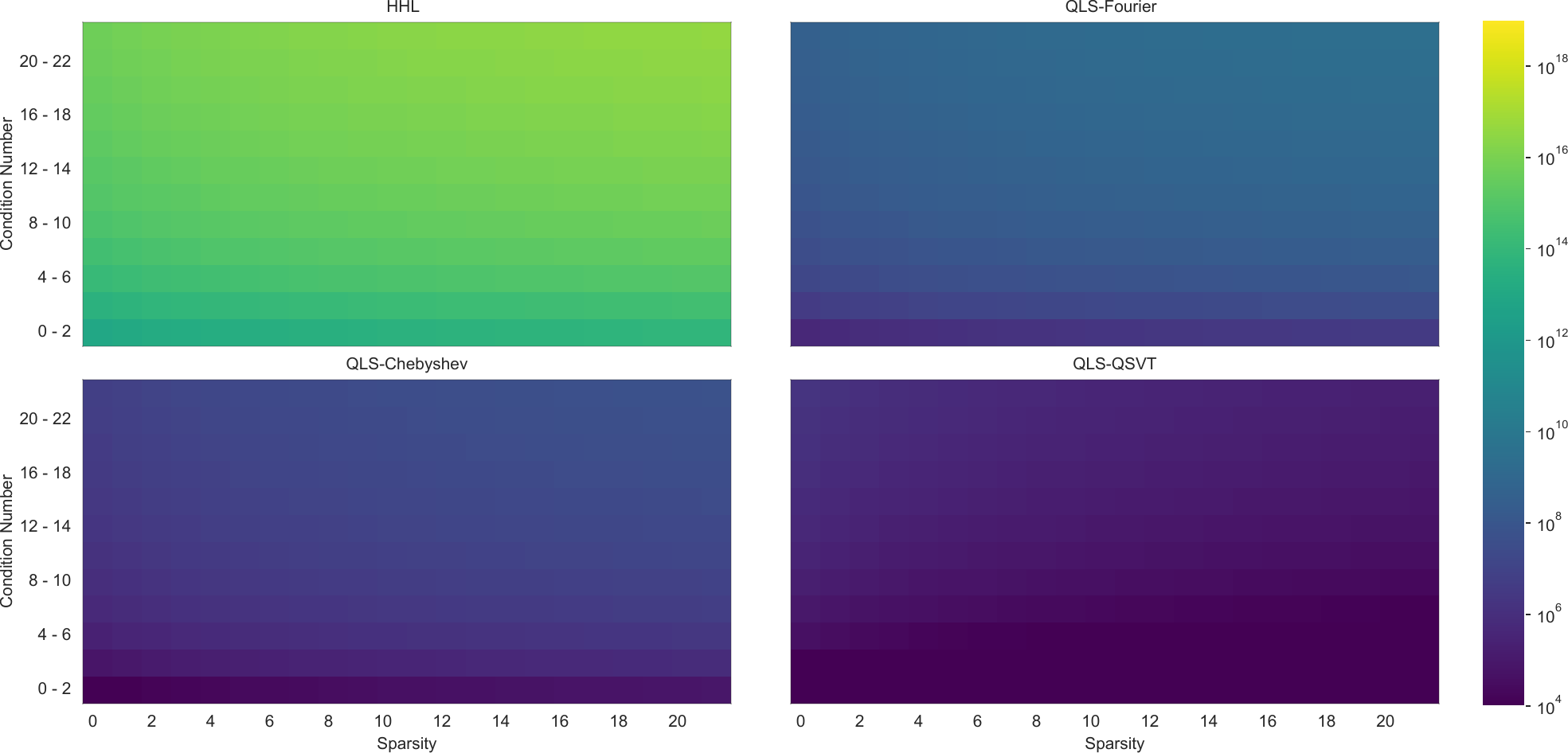}
    \caption{Condition number and sparsity heatmaps showing the average query count of the QLS. HHL, QLS-Fourier, QLS-Chebyshev, and QLS-QSVT query counts are given for the linear systems generated using a self-generated semi-random benchmark set. The color bar indicates the average number of queries required by each method. The color map illustrates the dependence on both the condition number and sparsity of each instance.}
    \label{fig:RandomGenerated}
\end{figure*}

\begin{restatable}[Query complexity of QLS-Chebyshev]{lemmma}{LemmaCostQLSC}\label{lemma:CostQLS-Chebyshev}
     Given access to the oracles in \eqref{equation:SparseOracles} specifying a $d$-sparse Hamiltonian $A$, the algorithm QLS-Chebyshev of \cite{Childs2017QuantumAlgorithmSistemyLinearEquations} prepares the state $A^{-1}\ket{b}$ up to normalization and with an accuracy of $\varepsilon$ using
    \begin{align*}
    \cost{[\qlsC]}&= 8j_0\nqaa
\end{align*}
queries to the oracles $\mathcal{O}_F$ and $\mathcal{O}_A$. 
The QAA overhead $\nqaa$ is computed using Eq.\ \eqref{n_q}, using the following expressions for the probability of success $p$ and its lower bound $p_0$
\begin{equation*}
    p_0= \frac{1}{\alpha^2} \, , \quad 
    p = \frac{\norm{x}^2}{\alpha^2} \, .
\end{equation*}
Here, $x=A^{-1}b$ is the solution of the linear system of equations, and 
\begin{align*}
    \alpha&=\frac{4}{d 2^{2s}}\sum_{j=0}^{j_0} {\sum_{i=j+1}^s{2s \choose s+i}} \, , \nonumber
    s=\lceil(d\kappa)^2\log_2(d\kappa/\varepsilon)\rceil \, , \\
    j_0&= \left\lceil \sqrt{\left\lceil d^2\kappa^2\log_2(d\kappa/\varepsilon)\right\rceil\log_2\left[\frac{4}{\epsilon}\left\lceil d^2\kappa^2\log_2(d\kappa/\varepsilon)\right\rceil\right]} \right\rceil \nonumber.
\end{align*}
\end{restatable}

\begin{restatable}[Query complexity of QLS-QSVT]{lemmma}{LemmaCostQLSQSVT}
   Given access to the oracles in \eqref{equation:SparseOracles} specifying a $d$-sparse Hamiltonian $A$, the algorithm of \cite{Gilyen2019QuantumSingularValueTransformation} prepares the state $A^{-1}\ket{b}$ up to normalization and with an accuracy of $\varepsilon$ using
\begin{equation*}
\begin{split}
\cost{[\qlsQ]}&=4n_{\text{rect}}\left(\frac{1}{d\kappa}, \: \min \left(\frac{\varepsilon}{4\kappa}, \: \frac{d\kappa}{2j_0(d\kappa, \frac{\varepsilon}{4\kappa})} \right)\right)\\
&+4n_{1/x}\left( {d\kappa} , \: \frac{\varepsilon}{4\kappa}\right) 
\end{split}
\end{equation*}    
queries to the oracles $\mathcal{O}_F$ and $\mathcal{O}_A$, where
\begin{equation*}
\begin{split}
    &n_{\text{rect}}\left(K, \varepsilon\right) =2n_{\text{exp}}\left(2k^2, \:  \frac{\sqrt{\pi}\varepsilon}{16k}\right)+1  \, , \\
    &n_{\text{exp}}\left(\beta, \varepsilon\right) = \left\lceil\sqrt{2\log(\frac{4}{\varepsilon})\left\lceil \max\left(\beta e^2 , \: \log(\frac{2}{\varepsilon})\right) \right\rceil} \right\rceil  \, , \\
    &k =\frac{\sqrt{2}}{K}\log^{1/2}\left(\frac{8}{\pi\varepsilon^2}\right)  \, ,
    n_{1/x} =2j_0 \left(\kappa, \: d , \: \frac{d}{2\kappa}\varepsilon \right)+1  \, .
\end{split}  
\end{equation*}
The QAA overhead $\nqaa$ is computed using Eq.\ \eqref{n_q}, using the following expressions for the probability of success $p$ and its lower bound $p_0$:
\begin{equation*}
    p =\frac{\norm{x}^2}{4\kappa^2} \, , \quad 
    p_0 =\left(\frac{1-\frac{\varepsilon}{2}}{2\kappa }\right)^2 \, .
\end{equation*}
\label{lemma:CostQSVTMI}
\end{restatable}


\section{Applications}\label{section:applications}
\begin{figure*}
    \centering
    \includegraphics[scale=0.55]{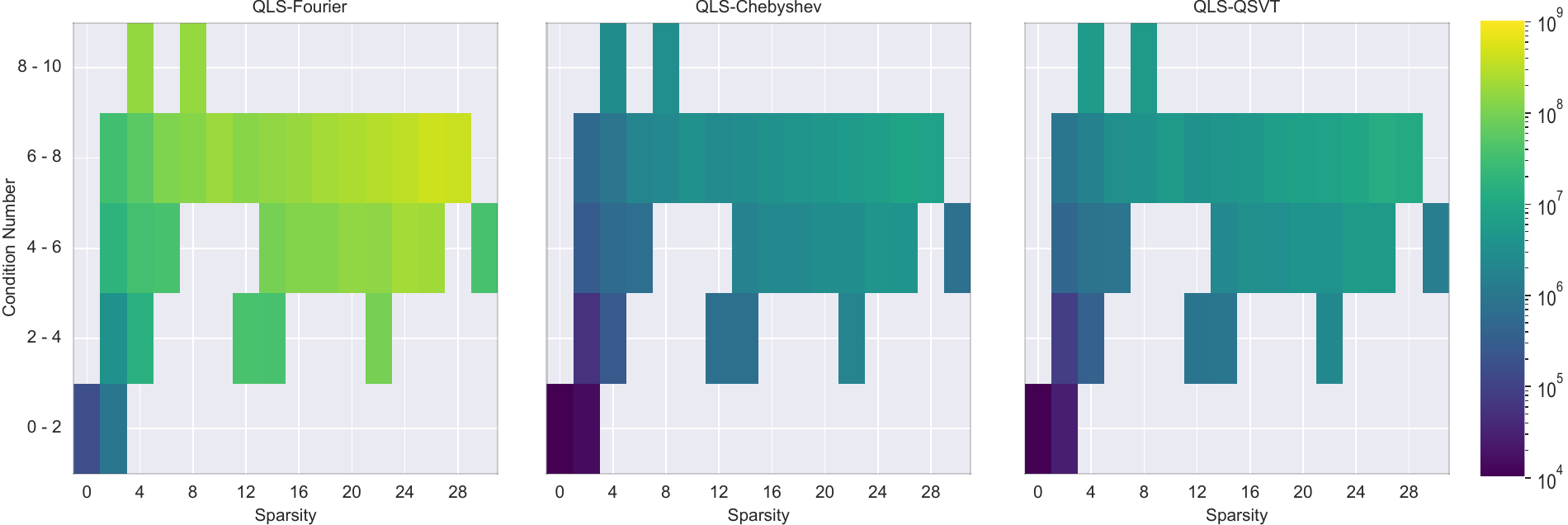}
    \caption{Query count heatmaps for QLS-Fourier, QLS-Chebyshev, and QLS-QSVT for the linear systems from simplex iterations. The color gradient shows the average number of queries required by each method, illustrating the joint dependence on condition number and sparsity.}
   \label{fig:Simplex}
\end{figure*}

QLS are used as a subroutine in several quantum algorithms. 
A non-exhaustive list includes differential equations \cite{leyton2008quantum, berry2014high, childs2021high}, data fitting \cite{wiebe2012quantum, schuld2016prediction, liu2017fast}, quantum machine learning \cite{rebentrost2014quantum}, computation of Green's functions \cite{tong2021fast}, and linear programming \cite{Nannicini2022FastQuantumSubroutinesSimplex}.
We refer the reader to the recent survey of quantum linear solvers \cite{morales2024quantum} for more details on applications.

To demonstrate the usefulness of our results, we focus on two possible applications here: linear programming, and partial differential equations (PDEs) -- in particular, Poisson equations.
We now briefly describe these problems and how they can be tackled using quantum linear solvers.

\subsection{Linear programming}
The goal of a linear program is to optimize a linear function subject to a linear constraint.
Namely, find $\min_x c^Tx$ subject to $Ax=b$ and $x\ge 0$, where $x,c\in \mathbb{R}^{N}$, $A\in\mathbb{R}^{M\times N}$, and $b\in \mathbb{R}^M$.
The most popular classical algorithm for solving linear programs is the simplex algorithm \cite{danzig1948linear}.
Geometrically, the algorithm can be viewed as a walk along the vertices of the convex polytope defined by the linear constraints. 
At each step of the walk, one needs to invert a matrix to decide in which direction to move: this accounts for most of the complexity of the algorithm.
In practice, the matrix is not actually inverted explicitly in each iteration, but rather the solver constantly updates an LU decomposition.

Recently, Nannicini has introduced a quantum version of the simplex algorithm \cite{Nannicini2022FastQuantumSubroutinesSimplex} which displays an asymptotic polynomial advantage over the classical algorithm. 
The algorithm is quite involved, and we refer to the original paper \cite{Nannicini2022FastQuantumSubroutinesSimplex} for all the details.
Here, it is important to note that the matrix inversion step of the simplex algorithm is replaced by a QLS.

The analysis of \cite{Ammann2023RealisticRuntimeAnalysisQuantumSimplex} shows that, using current state-of-the-art quantum subroutines, the asymptotic quantum advantage proven by \cite{Nannicini2022FastQuantumSubroutinesSimplex} does not apply to realistic instances of the problem. 
However, there is the possibility that future improvements of the algorithm from \cite{Nannicini2022FastQuantumSubroutinesSimplex}, either at the level of single subroutines or at the level of the algorithm as a whole, might reduce the gap to current classical solvers and eventually bring the theoretical advantage to instances of practical interest. 
For this reason, we still think that it is interesting to benchmark the different QLS considered in this work on linear systems of equations encountered while solving linear programs from instances of practical interest. 

\subsection{Partial differential equations}
As a model problem, we consider the Poisson equation with a Dirichlet boundary condition
\begin{equation*}
  \left\{
        \begin{aligned}
          -\Delta u &= f \quad \text{in } \Omega \, , \\
          u &= g \quad \text{on } \partial \Omega
        \end{aligned}
      \right.
\end{equation*}
with the $D$-dimensional hypercube $\Omega =  [-1, 1]^D$, $f \equiv 1$ and $g \equiv 0$.
The Poisson problem serves as a prototype for a second-order elliptic partial differential equation (PDE).
To obtain the data sets, we discretize the problem, resulting in a linear system $A \Tilde{u}_d = f_d$.
Here, $f_d, u_d, \Tilde{u}_d \in \mathbb{R}^N$ denote discretizations of $f$, $u$ and an approximation to $u$, respectively.
The matrix $A \in \R^{N \times N}$ captures the discretized derivatives.
Given $\varepsilon_d > 0$, the discretization had to be chosen such that $\norm{u_d - \Tilde{u}_d}_2 < \varepsilon_d$.
Since we assumed that the given PDE is second-order elliptic, $A$ is sparse, which motivates the application of a QLS.
Also note that we do not use a specialized quantum algorithm for this problem but instead consider that there exists a sparse access oracle that computes entries of the system matrix $A$ on demand.

\section{\label{section:Results}Results}
\begin{figure*}
    \centering
    \includegraphics[scale=0.54]{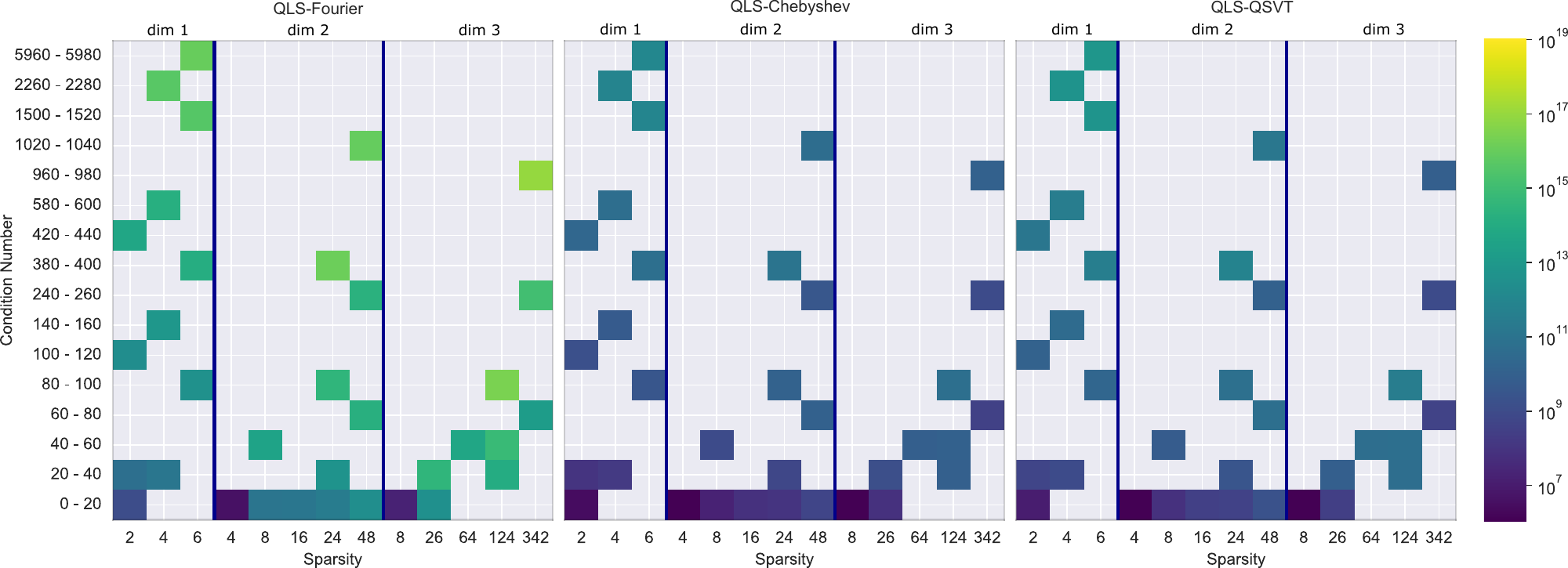}
    \caption{Comparison between the query count for QLS-Fourier, QLS-Chebyshev, and QLS-QSVT for the linear systems generated using 1-, 2- and 3-dimensional Poisson equations. The color bar indicates the average number of queries required by each method. This scaling is depicted as a color map, illustrating the dependence on both the condition number and sparsity of each instance.}
   \label{fig:Poisson}
\end{figure*}

We evaluate the performance of the functional quantum linear solvers on a broad set of benchmark instances.
The code and data are made publicly available.
\footnote{\url{https://gitlab.ibr.cs.tu-bs.de/alg/qls-comparison}}
We collect and generate a data set comprising instances from the following three classes:

A) \textbf{random}:
We perform the Hermitian matrix generation using a semi-random algorithm, which guarantees the normalization of the resulting matrix. The final data set consists of \num{1056000} problems with sizes ranging from \num{2} to \num{16777216} and was generated with the following steps: 
    \begin{enumerate}
        \item Randomly select $N_U$ values for $b_i \in [1, d]$ such that $\sum_{i = 1}^{N_U} b_i = N$ and at least one value, $b_j$, equals $d$. $N_U$ depends on the specific choice of the $b_i$.
        \item For each $i = 1, ..., N_U$, generate $U_i$, random unitary matrices of size $b_i \times b_i$. The unitary matrices are then arranged in a diagonal block matrix $U$. 
        \item Generate a random diagonal matrix $D = \text{diag}(\lambda_1, ..., \lambda_N)$, where the eigenvalues are constrained such that the maximum eigenvalue $\lambda_{\text{max}} = 1$, and the minimum eigenvalue $\lambda_{\text{min}} = 1/\kappa$, with $\lambda_{\text{min}} < \lambda_i < \lambda_{\text{max}}$ for all $i$. This ensures that the condition number of the resulting matrix is $\kappa$.
        \item $A = U D U^\dagger$ is a Hermitian matrix with the same condition number $\kappa$ and ensures that each row contains at most $d$ non-zero values.
        \item Apply a random permutation to $A$. 
        \item Generate vector $b$ as a normalized random complex vector of dimension $N$.
    \end{enumerate}
    
B) \textbf{simplex}:
We generate a set of linear systems from simplex iterations on MIPLIB~\cite{miplib2021} instances. Specifically, we run the GNU Linear Programming Kit (GLPK) simplex solver, extracting systems from iterations where the current basis is well-conditioned. Since computing the condition number of the basis matrix in each iteration is computationally expensive, we restrict our selection to reasonably small instances. 
The final data set consists of \num{106} LP problems with basis sizes ranging from \num{5} to \num{223082}, which produces a total of \num{10233} well-conditioned iterations.

C) \textbf{Poisson}:
We use the deal-ii C++ library \cite{deal-ii} to build linear systems, allowing us to use dimension-independent programming to easily incorporate different values of $D$. The benchmarking code is based on the deal-ii tutorial \verb|step-4|~\cite{PDEsGenerator} which uses a finite element ansatz for discretization. It has been modified to generate data for $D \in \{1, 2, 3\}$, the number of grid refinements $n_r \in [1, 5]$, and the degree $d_{\text{FE}} \in [1, 3]$ of the finite element polynomials.
The final data set consists of \num{90} problems with sizes ranging from \num{2} to \num{117649}.
The accuracy value chosen for all benchmark results is $\varepsilon = 10^{-8}$. 

In \autoref{fig:RandomGenerated}, we compare the query count for HHL, QLS-Fourier, QLS-Chebyshev, and QLS-QSVT for the linear systems generated using the \textbf{random} benchmark set. 
Our results show a significant gap in the number of query calls: HHL requires between $10^{12}$ and $10^{18}$ queries, while the other methods require between $10^{4}$ and $10^{10}$ queries. 
A reason for this gap is that the dependence of HHL on epsilon, which is $1/\varepsilon$, compared to $\log(1/\varepsilon)$ for the other methods, has a practical computational disadvantage. 
We can then exclude HHL query counts from our subsequent query comparisons. 
Also note that the scaling of query calls for QLS-QSVT with respect to the same parameters remains below $10^8$, showing its better efficiency in comparison with the other three methods.

In \autoref{fig:Simplex} and \autoref{fig:Poisson}, we compare the query counts for QLS-Fourier, QLS-Chebyshev, and QLS-QSVT for the linear systems generated using the \textbf{simplex} and \textbf{Poisson} data sets, respectively. 
In both cases, QLS-Chebyshev and QLS-QSVT consistently outperform QLS-Fourier, with query counts remaining below $10^{8}$ for the simplex set and between $10^{6}$ and $10^{14}$ for the Poisson set. 
In contrast, QLS-Fourier requires query counts ranging from $10^{4}$ to $10^{10}$ for the simplex set, and from $10^{10}$ to $10^{15}$ for the Poisson set. 


\section{\label{section:DiscussionAndConclusion}Discussion and Conclusions}
In this work, we compared the query counts of the functional quantum linear solvers HHL, QLS-Fourier, QLS-Chebyshev, and QLS-QSVT.
We employed a hybrid benchmarking technique by deriving estimates on the numbers of calls to the matrix oracles $\mathcal{O}_F$ and $\mathcal{O}_A$, while classically recording relevant problem parameters.

The evaluation was conducted across a broad set of benchmark instances.
These included a self-generated semi-random benchmark set specifically designed to satisfy the ideal conditions for QLS, allowing us to assess the potential of each algorithm under optimal settings.
Additionally, we included two real-world problem instance collections: linear systems derived from simplex iterations within MIPLIB and partial differential equations (Poisson equations).

Our motivation was to quantify which quantum linear solver has the smallest estimated query costs, across several relevant real-world problems. In our analysis, the Chebyshev 
polynomial-based methods (QLS-Chebyshev and QLS-QSVT) have the smallest estimated query count for the real-world problems that we tested. 
Moreover, while HHL remains historically and theoretically important, its query count is too high to be considered practical for most real-world applications, especially due to its underperformance on the instances tailored to provide optimal conditions for all QLS. 
The gap in performance between HHL and QLS-Fourier underscores that we would not expect this result to change with a different choice of Hamiltonian simulation.

Finally, notice that it would be interesting to extend this comparison to other QLS algorithms, in particular to the class of so-called adiabatic QLS. These algorithms use  methods closely related to adiabatic quantum computing \cite{Subasi2019QuantumAlgorithmSystemsLinearEquationsAdiabaticQuantumComputing} to create a QLS. Combined with an eigenvalue filtering routine from \cite{Dong2022QuantumLinearSysteSolverAdiabationQuantumComputingQAOA}, this approach has led to a quantum algorithm with optimal scaling in the accuracy and condition number \cite{Costa2022OptimalScalingQuantumLinearSystemsSolverDiscreteAdiabaticTheorem}. One challenge in doing this is that adiabatic QLS are structurally distinct from functional QLS, and therefore more difficult to compare with the same assumptions.

\section*{Acknowledgment}
Helpful discussions with Paul Herringer, Tobias Osborne and Thomas Wick are gratefully acknowledged.
This work was supported by the Quantum Valley Lower Saxony, the BMBF project QuBRA, and the BMWi project ProvideQ. M.S. gratefully acknowledges the Deutsche Forschungsgemeinschaft (DFG) under Germany’s Excellence Strategy within the Cluster of Excellence PhoenixD (EXC 2122, Project ID 390833453). S.E. Skelton acknowledges the support of the Natural Sciences and Engineering Research Council of Canada (NSERC), PGS D - 587455 - 2024. Cette recherche a été financée par le Conseil de recherches en sciences naturelles et en génie du Canada (CRSNG), PGS D - 587455 - 2024.

\section*{Author contributions}
This project was initiated and constructed in and from the discussions of A.I.L., D.R., A.F.R, and S.E.S. The benchmark instances for the Poisson equations were provided by M.S., and the linear programming ones by M.P. All authors contributed to writing the paper.

\section*{Notation cheat sheet}
\begin{table}[h!]
    \begin{center}
    \bgroup
\def\arraystretch{1.2} 
\begin{tabular}{ c | c }
 Notation & Interpretation\\ 
 \hline
$A \in \mathbb{C}^{N\times N}$ & $N \times N$ invertible Hermitian matrix \\ 
$\norm{A}_{\text{max}}$ & value of the largest entry in $A$ \\
$b \in \mathbb{C}^N$ & constraints vector \\
$x \in \mathbb{C}^N$ & solution vector \\
$d$ & maximum sparsity per row or column, \\
$\kappa$ & condition number, \\
$\dom{\kappa}$ & $[-1, -1/\kappa] \cup [1/\kappa, 1]$\\
$\mathcal{O}_F$ & $\mathcal{O}_F\ket{j, l} = \ket{j, f(j, l)}$\\
$\mathcal{O}_A$ & $\mathcal{O}_A\ket{j, k, z} = \ket{j, k, z\oplus A_{jk}}$\\
$p$ & probability of finding a good state, \\
& $p = \sin^2{\theta}$ \\
$p_0$ & lower bound on $p$ \\
$t$ & Hamiltonian simulation time \\
$\nqaa$ & expected number of queries to QAA \\
$\varepsilon$ & solution accuracy \\
\hline
\end{tabular}
\egroup
\end{center}
\label{tab:notation}
\end{table}

\bibliographystyle{unsrt}
\bibliography{main.bib}

\newpage
\appendices

\section{\label{section:AppendixHamiltonianSimulation}Hamiltonian Simulation - Qubitization}
In this section, we sketch the basic framework of Hamiltonian simulation by qubitization as introduced by Low and Chuang in \cite{Low2019Qubitization}. One way to understand QSP is as an operation which applies a sum of Chebyshev polynomials on a quantum circuit. This is because polynomials on $x\in [-1, 1]$ can be written as sums of Chebyshev polynomials through the coordinate transformation $x=\cos(\theta)$.\\

The goal is to efficiently approximate the unitary evolution $e^{-iAt}$ of a $d$-sparse matrix $A$.
Hamiltonian simulation in \cite{Low2017OptimalHamiltonianSimulationQuantumSignalProcessing} uses a functional approximation for $e^{iAt}$, dependent on $\varepsilon, t$.
Implementing this with quantum signal processing further requires an oracle which prepares the entries of $A$, depending on the $n$ system qubits in the oracle and a couple of ancillary qubits.

\LemmaCostQubitization*

\begin{proof}
    We can use Lemma \ref{lemma:encodingdsparsehamiltonians} to create a block encoding. As in \cite{Low2019Qubitization}, the block encoding requires two uses of $T$, so the block encoding uses a total of $6$ calls to $\mathcal{O}_F, \mathcal{O}_A$. We then build the qubitized operator $\Tilde{W}$ with one call to the block-encoding and its inverse\cite{Low2019Qubitization}, leading to an oracle with $12$ calls to $\mathcal{O}_F, \mathcal{O}_A$. Then the quantum eigenvalue transformation (the QSP extension for qubitized oracles) is used to approximate $\exp\left({it\frac{A}{d||A||_{\text{max}}}}\right)$ within $\varepsilon$. Finally, we introduce the coordinate change $\Tilde{t}= d||A||_{\text{max}}t$ so that the overall approximation is $\varepsilon$ close to $\exp\left({it{A}}\right)$.\\
    
    The costs of the QSP procedure are determined by the polynomial degree of the approximation. We compute this using bound $\Tilde{r}$ in Lemma \ref{lemma:BoundOnrQSVT}, with arguments $\Tilde{t}$ and $\varepsilon$. The functional approximation Lemma \ref{lemma:BoundOnrQSVT} is of the correct form to be used in Lemma \ref{lemma:QSP}, where $\frac{N}{2}=2\Tilde{r}$ and the QSP procedure needs $N$ calls to $\Tilde{W}$. So finally we are justified in taking the query complexity as 
    \begin{equation}
        \cost{[e^{-iAt}]}=48\Tilde{r}\left(\Tilde{t}, \varepsilon\right) \, .
    \end{equation}
\end{proof}

\subsection{Quantum Signal Processing}
Quantum signal processing is a general routine for applying some 'suitable' polynomial to the eigenvalues of unitary $U_A$. 
There are many QSP conventions - \cite{Liu2021RigorousRobustQuantumSpeedUpSupervisedMachineLearning, Gilyen2019QuantumSingularValueTransformation} can be generalized to QSVT, or presented with a potentially multiplicatively optimal circuit such as in \cite{Berry2024DoublingEfficiencyHamiltonianSimulationQSVT, Motlagh2024GeneralizedQuantumSignalProcessing}, but we will use the presentation for Hamiltonian simulation. First we recall the main QSP result Lemma \ref{lemma:QSP} which formalizes how QSP operator $U_{h}$ encodes polynomial $A+iB$ applied to the eigenvalues of $W$.
\begin{lemma}[QSP from \cite{Low2017OptimalHamiltonianSimulationQuantumSignalProcessing}, modified]\label{lemma:QSP}
    Assume a unitary $W$, defined in its basis of eigenvalues $e^{i\theta}$ by $W(\theta)=\sum_{\theta}e^{i\theta}\ket{\theta}\bra{\theta}$.
    Let ${h: (-\pi, \pi]\rightarrow (-\pi, \pi]}$ be a real odd periodic function, $N>0$ even, and let $[A(\theta), C(\theta)]$ be real Fourier series in $\left[\cos(k\theta), \sin(k\theta)\right]$, $k=0, 1, 2, \ldots \frac{N}{2}$, that approximate 
    \begin{equation}
    \max_{\theta\in\mathbb{R}}\left|A(\theta)+iC(\theta)-e^{ih(\theta)}\right|\leq \varepsilon \, .
    \end{equation}
Then there exists a set of angles $\Phi$ such that for the circuit defined by
\begin{align}
    U_{h}&=\prod_{j=1}^NU_{\phi_j}(\theta),\\
    U_{\phi_j}(\theta)&=\left(e^{-i\frac{\phi_j}{2}\sigma_Z}\otimes I\right)C_{+}W\left(e^{i\frac{\phi_j}{2}\sigma_Z}\otimes I\right),\\
    C_+W(\theta)&=\ket{+}\bra{+}\otimes I +\ket{-}\bra{-}\otimes W(\theta)\nonumber \, ,\\
\end{align}
we have that $\left|\bra{\theta}\bra{+}U_{h}\ket{+}\ket{\theta}-e^{ih(\theta)}\right|\leq c\varepsilon$ for any eigenvalue determined by $\theta$. 
\end{lemma}
The constant $c$ is determined by pre-processing method used, see for example
\cite{Haah2019ProductDecompositionPeriodicFunctionsQSP, skeltonupcoming2024}. Within this work, we will disregard the pre-processing loss of accuracy, that is we will assume $c=1$.

\subsection{Functional Approximation}
Recall the Jacobi-Anger expansion
\[
e^{i\cos (z) t}=\sum_{k=-\infty}^{\infty} i^k J_k(t) e^{i k z} \, ,
\]
where $J_k(t)$ are Bessel functions of the first kind \cite{AbramowitzMilton1974HandbookOfMathematicalFunctions}. We will work in variable $\cos(z)=\lambda\in [-1, 1]$. The polynomials $A$ and $C$ are obtained by truncating and re-scaling Jacobi-Anger expansion using the Fourier-Chebyshev series of the trigonometric functions
\begin{equation}
\begin{aligned}
   e^{-i\lambda t}&=\underbrace{J_0(t)+2\sum_{k=1}^{\infty}(-1)^{k}J_{2k}(t)T_{2k}(\lambda)}_{\cos{(t\lambda)}}\\
   &+i \underbrace{2\sum_{k=0}^{\infty}(-1)^{k}J_{2k+1}(t)T_{2k+1}(\lambda)}_{\sin(t \lambda)} . 
\end{aligned}
\end{equation}
The degree-$2R$ truncation is provably a polynomial with the necessary conditions for use in QSP. 
By following the reasoning in \cite{Gilyen2019QuantumSingularValueTransformation, Berry2015HamiltonianSimulationNearlyOptimalDependenceOnAllParameters}, we find a degree $R$ polynomial such that 
\begin{equation}
\begin{aligned}
  &\left|\left|\cos(t\lambda)-J_0(t)-\sum_{k=1}^R(-1)^kJ_{2k}(t)T_{2k}(\lambda)\right|\right|_{\infty} \leq \varepsilon\\
  &\left|\left|\sin(t\lambda)-\sum_{k=1}^R(-1)^kJ_{2k+1}(t)T_{2k+1}(\lambda)\right|\right|_{\infty} \leq\varepsilon \, ,
\end{aligned} 
\end{equation}
where $\left|\left|\cdot\right|\right|_{\infty} $ is the $l$-infinity norm over the relevant domain, in this case $\lambda\in \mathcal{D}_0 = [-1, 1]$.

For all $m\in \mathbb{N}_+$, and $t\in \mathbb{R}$ it holds that $\left|J_m(t)\right|\leq\frac{1}{m!}\left|\frac{t}{2}\right|^m$ \cite{AbramowitzMilton1974HandbookOfMathematicalFunctions}. We require $2\sum_{l=0}^{\infty}\left|J_{2R+2+2l}(t)\right|\leq \varepsilon$. Hence, for any positive integer $q\geq 2R+2$, and $q\geq t-1$, we obtain
\begin{equation}
\begin{aligned}
      2\sum_l^{\infty}\left|J_{q+2l}\right|&\leq 2\left|\frac{t}{2}\right|^{q}\sum_l^{\infty}\frac{\frac{1}{4^l}\left|{t}\right|^{2l}}{q!(q+1)(q+2)\dots(q+2l)}\\
      &\leq 2\left|\frac{t}{2}\right|^{q}\frac{1}{q!}\sum_l^{\infty}\left(\frac{1}{4}\right)^{l}
      \leq\frac{8}{3q!}\left|\frac{t}{2}\right|^{q}, 
\end{aligned}
\end{equation}
where we bounded the product of the $2l$ terms of the form $\frac{t}{(q+a)}, a=1\dots2l$ by one.

By using Stirling's inequality $q!\geq \sqrt{2\pi q}\left(\frac{q}{e}\right)^q$, we obtain
\begin{equation}
\begin{aligned}
     2\sum_l^{\infty}\left|J_{q+2l}\right| &\leq  \frac{8}{3\sqrt{2\pi q}}\left(\frac{e\left|t\right|}{2q}\right)^q\leq\frac{8}{3\sqrt{2\pi}}\frac{1}{\sqrt{2}}\left(\frac{e\left|t\right|}{2q}\right)^q\\
     &=\frac{4}{3\sqrt{\pi}}\left(\frac{et}{2q}\right)^q .
\end{aligned}
\end{equation}

Therefore, we have
\begin{align}
    \left|\left|\cos(t\lambda)-J_0(t)-\sum_k(-1)^kJ_{2k}(t)T_{2k}\right|\right|_{\infty} &\leq \varepsilon \, ,\\
    \left|\left|\sin(t\lambda)-\sum_k(-1)^kJ_{2k+1}(t)T_{2k+1}\right|\right|_{\infty} &\leq \varepsilon \, ,
\end{align}
for
\begin{equation}\label{eq;lambeth_thing_for_bound}
    \frac{4}{3\sqrt{\pi}}\left(\frac{et}{4(R+1)}\right)^{2(R+1)}\leq\varepsilon  \, .
\end{equation}
The next step is to identify $R$ as a function of $t>0$, and $\varepsilon$. 
We define  $r(t, \varepsilon)$ as a solution to
\begin{equation}
    \left(\frac{et}{2r}\right)^r=\varepsilon \, .
\end{equation}
This function is related to the W-Lambeth function and we have that
\begin{equation}
    2(R+1)=\frac{1}{2}\max\left(\left\lceil r\left(\frac{et}{2}, \varepsilon\right)\right\rceil, \left|t\right|\right)-1  \, .
\end{equation}
Unfortunately,  $r(t, \varepsilon)$ cannot always be solved.
Importantly, for $R_1, R_2\in[t, \infty)$, $\left(\frac{et}{2R_1}\right)^{R_1}\geq\varepsilon\geq \left(\frac{et}{2R_2}\right)^{R_2}$ implies $R_1\leq r(t, \varepsilon)\leq R_2$. 
In our case, $R_1=2(R+1)$, $\left(\frac{et}{2R_1}\right)^{R_1}=\varepsilon\geq \left(\frac{et}{2R_2}\right)^{R_2}$ and we want an upper bound $R_2$ on  $r(\frac{et}{2}, \varepsilon)$.

This leads to a precise version Lemma 59 of \cite{Gilyen2019QuantumSingularValueTransformation} (which itself follows \cite{Low2019Qubitization, Berry2015HamiltonianSimulationNearlyOptimalDependenceOnAllParameters}).

\begin{lemma}[Bound on $r(t, \varepsilon)$ \cite{Gilyen2019QuantumSingularValueTransformation}]
    For $t\in\mathbb{R}_+$ and $\varepsilon\in(0, 1)$, $\Tilde{r}(t, \varepsilon)$ is an upper bound on  $r(t, \varepsilon)$
    \begin{equation}
        \Tilde{r}(t, \varepsilon)=\begin{cases}
            \lceil et\rceil & t\geq\frac{\ln(1/\varepsilon)}{e}  \, ,\\
             \lceil \frac{4\log(1/\varepsilon)}{\log\left(e+\frac{1}{t}\log(1/\varepsilon)\right)}\rceil & t< \frac{\ln(1/\varepsilon)}{e} \, .
        \end{cases}
    \end{equation}
    Moreover, for all $q\in\mathbb{R}_+$,
    \begin{equation}
        r(t, \varepsilon)<e^qt+\frac{\ln(1/\varepsilon)}{q} \, .
    \end{equation}
    \label{lemma:BoundOnrQSVT}
\end{lemma}
We have modified the original Lemma 59 from \cite{Gilyen2019QuantumSingularValueTransformation} to obtain precise bounds for $r(t, \varepsilon)$, by using the following results

\textbf{Case 1.} For all $t\geq \frac{\ln(1/\varepsilon)}{e}$, $r(t, \varepsilon)\leq et$.

\textbf{Case 2.} For all  $t\leq \frac{\ln(1/\varepsilon)}{e}$, $\frac{\ln(1/\varepsilon)}{\ln(\left(e+\frac{1}{t}\ln(1/\varepsilon)\right))}\leq r(t, \varepsilon)\leq \frac{4\ln(1/\varepsilon)}{\ln(\left(e+\frac{1}{t}\ln(1/\varepsilon)\right))}$.

In the original proof, these terms are added together in order to obtain an asymptotic bound for all $t$. 
However, if we know which case we are in, then we can use the expression quoted in Lemma \ref{lemma:BoundOnrQSVT}. At $t= \frac{\ln(1/\varepsilon)}{e}$, both cases are valid but case $1$ provides a tighter bound.

\section{\label{section:AppendixQAA}Quantum Amplitude Amplification}
Given a boolean function $f:Z\rightarrow\{0,1\}$ partitioning the set $Z$ into a set of $\textit{good}$ elements $G = \{z \in Z\; | \; f(z) = 1 \}$ and one of $\textit{bad}$ elements $B =\{z \in Z \; | \; f(z) = 0 \}$, let $\mathcal{A}$ be a quantum algorithm preparing a good state with probability $p = \sin^2{\theta}$
\begin{equation}
    \mathcal{A}\ket{0} = \sin{\theta}\ket{\psi_{G}} + \cos{\theta}\ket{\psi_{B}}.
\end{equation}
Here, we have defined the projections of $\mathcal{A}\ket{0}$ on the good and bad subspaces
\begin{equation}
\ket{\psi_G}\coloneqq {\Pi}_G\cdot \mathcal{A}\ket{0} , \quad \ket{\psi_B}\coloneqq \Pi_B\cdot \mathcal{A}\ket{0} ,
\end{equation}
with $\Pi_G=\sum_{z\in G}\ketbra{z}$ and $\Pi_B=\sum_{z\in B}\ketbra{z}$.

The goal of quantum amplitude amplification (QAA) is to boost the probability of finding a good element upon measurement. 
This is achieved by repeatedly applying the following operator
\begin{align}\label{equation:AmplitudeAmplificationOperator}
    Q \coloneqq \mathcal{A} \mathcal{S}_{0} \mathcal{A}^{\dagger} \mathcal{S}_{f} \, ,
\end{align}
where $\mathcal{S}_{0}$ changes the sign of the amplitude of the state $\ket{0}$, and $\mathcal{S}_{f}$ is the \emph{phase oracle} that flips the sign of the amplitudes
\begin{equation}\label{equation:PhaseOracle}
    \mathcal{S}_{f} \ket{z} \coloneqq (-1)^{f(z)} \ket{z}.
\end{equation}

When the probability of success $p$ is known, \cite{Brassard2002QuantumAmplitudeAmplificationEstimation} showed that one should apply $Q$ $m=\lfloor \pi/(4\arcsin \sqrt{p})\rfloor$ times. 
The number of times we apply $\mathcal{A}$ is then $2m$. When only a lower bound on the probability of success $p_0\le p$ is known, one has to use a randomized version of the algorithm where $Q$ is applied a random number of times sampled from a growing interval.
In all the algorithms considered in this paper, only a lower bound on the probability of success is known, so we consider this version of the algorithm.
We summarize it in Algorithm \ref{alg:qaa}.

\begin{algorithm}
\caption{QAA}\label{alg:qaa}
\begin{algorithmic}[1]
\Function{QAA}{$\mathcal{A}$, $f$}
\State Set $k = 0$, and $1 < c < 2$\;
\State \textbf{while} True \textbf{do}
\State Increase $k$ by 1, $m_k = \lfloor\min(c^k, \sqrt{1/p_0})\rfloor$\;
\State Apply $\mathcal{A}$ to $\ket{0}$\;
\State Choose $j \in [0, m_k]$ uniformly at random\;
\State Apply $Q^{j}$ to the register\;
\State Measure the register, let $z$ be the outcome\;
\State \textbf{if } $f(z) = 1$
\State  return $z$\;
\EndFunction
\end{algorithmic}
\end{algorithm}

Given that the algorithm is randomized, one cannot know exactly how many times $\mathcal{A}$ is applied in total before a good state is found. 
This number scales like $\mathcal{O}(1/\sqrt{p})$ \cite{Brassard2002QuantumAmplitudeAmplificationEstimation},  as in the case in which the probability of success is known.
However, to benchmark the algorithms we need a better estimate for it.
In the Lemma below, we provide a formula for the expected number of calls to $\mathcal{A}$ that the quantum algorithm need to find a good state. 
To compute this formula we need to classically compute the exact probability of success $p$.
Notice that we \emph{don't} assume that this is known when running the quantum algorithm, but we only compute it to perform the benchmarking.

\LemmaCostQAA*

\begin{proof}
Let 
\begin{equation}
\ket{\psi_0}\coloneqq \mathcal{A}\ket{0}=\sin \theta\ket{\psi_G}+\cos \theta\ket{\psi_B},
\end{equation}
where $\sin ^2 \theta= p$. 
As shown in \cite{Brassard2002QuantumAmplitudeAmplificationEstimation}, 
the operator $Q$ from Eq.\ \ref{equation:AmplitudeAmplificationOperator} generates translations in $\theta$. 
Let $\left|\psi_j\right\rangle=Q^j\left|\psi_0\right\rangle$, then one can show that
\begin{equation}
    \left|\psi_j\right\rangle=\sin [(2 j+1) \theta]\left|\psi_G\right\rangle+\cos [(2 j+1) \theta]\left|\psi_B\right\rangle .
\end{equation}
This implies that, after $j$ applications of $Q$, the probability of finding a good state as the outcome of the measurement is
\begin{equation}\label{eq:pj}
  p_j=\sin ^2[(2 j+1) \theta] .  
\end{equation}
At each iteration of the while loop in Algorithm \ref{alg:qaa}, $Q$ is applied $j$ times (see step 7), with $j$ sampled uniformly at random in $\{0,... m_k\}$ and $m_k = \lfloor\min(c^k, \sqrt{1/p_0})\rfloor$.
The measurement is then successful with probability $p_j$ as in Eq.\ \ref{eq:pj}.
In algorithm \ref{alg:qaa} we begin by sampling $j_1$ and applying $Q$ $j_1$ times. With probability $1 - p_{j_1}$ the algorithm is not successful, and we need to sample $j_2$, and apply $Q$ additional $j_2$ times.
Similarly, with probability $1 - p_{j_2}$ the algorithm doesn't succeed, hence we sample $j_3$ and so on.
Let $J=(j_1, j_2, j_3\dots )$ be the $j_k$ sampled this way, then the expected number of applications of $Q$ given $J$ is
\begin{equation} \label{eq:AvgNumberApplicationsQ}
\begin{split}
\left\langle n_Q\right\rangle_{J}&= j_1 + j_2(1-p_{j_1})+ j_3(1-p_{j_1})(1-p_{j_2})+\dots \\
&=\sum_{k=1}^{\infty}\left[j_k \prod_{l=1}^{k-1}\left(1-p_{j_l}\right)\right] \, .
\end{split}
\end{equation}
We then to need to average over the randomly chosen $j$'s.
One finds 
\begin{equation}
  \langle n_Q\rangle =\sum_{k=1}^{\infty}\left[\left\langle j_k\right\rangle \prod_{l=1}^{k-1}\left(1-\left\langle p_{j_l}\right\rangle\right)\right],  
\end{equation}
with 
\begin{equation}
\begin{aligned}
\left\langle p_{j_k}\right\rangle & =\frac{1}{m_k} \sum_{j_k=0}^{m_k} p_{j_k} \\
& =\frac{1}{2}-\frac{\sin(4(m_k + 1)\theta)}{4(m_k + 1) \sin (2 \theta)}  \, .
\end{aligned}  
\end{equation}
and 
\begin{equation}
 \left\langle j_k\right\rangle=\frac{1}{m_k+1} \sum_{j_k=0}^{m_k} j_k=\frac{m_k}{2}  \, .
\end{equation}
Finally, notice that each application of $Q$ requires two applications of $\mathcal{A}$ (or its inverse). Hence, we arrive at Eq.\ \eqref{n_q}. 
\end{proof}

\section{\label{section:AppendixHHL}HHL}
In \cite{Harrow2009QuantumAlgorithmLinearSystemsEquations}, Harrow, Hassidim, and Lloyd introduced the first QLS algorithm, which after their initials is now called the HHL algorithm. 
This algorithm has since been improved upon by many newer QLS algorithms, and we don't expect it to be competitive. 
Nonetheless, it remains one of the best-known QLS algorithms, and hence we include it in our comparison.

Let's first describe the algorithm intuitively for a very special class of linear systems.
Let $\ket{u_j}$ be the eigenstates of $A$, and $\lambda_j$ the corresponding eigenvalues, with $j=1,\dots, N$: we assume $t_0\lambda_j=2\pi \tau_j$, for a suitable $j$-independent constant $t_0\in \mathds{R}$, and suitable integers $\tau_j\in \{0, \ldots, T-1\}$. 
Notice that by assumption $\lambda_j\in[1/\kappa,1]$,
from which it follows that $\tau_j\ge t_0/(2\pi\kappa)$ and $\tau_j\le t_0/(2\pi)$. 
From this second bound, it follows that it is sufficient to take $T=1+\lfloor t_0/(2\pi) \rfloor$.

The algorithm starts by preparing the state $\ket{b}$ using the oracle $\mathcal{P}_b$ and then applying QPE with accuracy $T$ and the unitary $U=\exp(iAt_0/T)$. 
This is done in three steps. 
First, we prepare an ancillary register of dimension $T$ in a uniform superposition, $\sum_{\tau=0}^{T-1}\ket{\tau}$.
We will call this register the \emph{clock}. 
We then apply the following controlled unitary
\begin{equation}\label{eq:hhl_U}
    \mathcal{U}=\sum_{\tau=0}^{T-1}\ketbra{\tau}_c\otimes \exp{iA\frac{t_0}{T}\tau}\,, 
\end{equation}
where the subscript $c$ indicates that the unitary is controlled on the clock register. 
Finally, we apply the inverse quantum Fourier transform.
A simple calculation shows that the resulting state is $\sum_{j=1}^{N}\beta_j\ket{\tau_j}\otimes\ket{u_j}$, where $\beta_j$ are the components of $\ket{b}$ in the eigenbasis of $A$.

Notice that the eigenvalues of $A$ are encoded in the first register of the state above, $\ket{\tau_j}$, through $\lambda_j = 2\pi \tau_j/t_0$. 
In the eigenbasis of $A$, we can implement the inverse of $A$ by simply inverting the eigenvalues. 
Given the relation between $\lambda_j$ and $\tau_j$, and the fact that the inverse is computed up to an overall constant, it is enough to transform the first register as
\begin{equation}\label{eq:invert_j}
    \ket{\tau}\rightarrow \frac{C}{\tau}\ket{\tau}\,
\end{equation}
with a normalization constant $C$.
Note that this transformation is $j$-independent, so we do not need information on the spectrum of $A$ to implement it.  
However, it is not unitary and cannot be implemented directly. 

The authors of \cite{Harrow2009QuantumAlgorithmLinearSystemsEquations} implement \eqref{eq:invert_j} as follows. They introduce an additional ancilla qubit prepared in the state $\ket{0}$, and rotate it conditioned on the state of the clock register with\footnote{Here,
\begin{equation}
    R_Y(\theta) \coloneqq \begin{pmatrix}
                \cos(\theta/2) & -\sin(\theta/2) \\
                \sin(\theta/2) & \cos(\theta/2)
                \end{pmatrix}\,.
\end{equation}} 
\begin{equation}\label{eq:hhl_R}
    \mathcal{R} =\sum_{\tau=0}^{T-1}\ketbra{\tau}_c\otimes R_Y(\theta_\tau)\,,
\end{equation}
where $\theta_\tau$ satisfies
\begin{equation}\label{eq:theta_tau}
    \sin\frac{\theta_\tau}{2}=\frac{C}{\tau}\,, \text{ for } \tau\ge \frac{t_0}{4\pi\kappa}\,,
\end{equation}
and $\theta_\tau=0$ otherwise.
The reason why we can set $\theta_\tau=0$ for small values of $\tau$ is that $\lambda_j\ge 1/\kappa$. The constant $C$ must be chosen such that the r.h.s.\ of \eqref{eq:theta_tau} is smaller than one: it is enough to take $C=t_0/(4\pi\kappa)$. To simplify the notation, let $\tilde{f}_\tau \coloneqq \sin(\theta_\tau/2)$.
After this rotation, we undo the QPE and trace away the clock register.
This yields
\begin{equation}
    \sum_{j=0}^{T-1}\beta_j\ket{u_j}\otimes \Bigl(\sqrt{1-\tilde{f}_\tau^2} \ket{0}+\tilde{f}_\tau \ket{1}\Bigr)\,.
\end{equation}
Finally, we measure the flag ancilla: if the result is 1, the rest of the state collapses to $\ket{x}$; if the result is 0, we know that the algorithm has failed. 
For this reason, we call this ancilla the \textit{success flag}. 
The success probability is given by 
\begin{equation}\label{eq:probFlagSimple}
    p=\frac{1}{16\pi^2}\frac{t_0^2}{\kappa^2}\sum_{j=1}^{N}\frac{\abs{\beta_j}^2}{\tau_j^2}\,.
\end{equation}
We can lower bound this expression recalling that $\tau_j\le t_0/(2\pi)$, leading to 
\begin{equation}\label{eq:low_bound_hhl_p_1}
    p\ge \frac{1}{4\kappa^2}\,.
\end{equation}
We can boost this probability to $\mathcal{O}(1)$ using quantum amplitude amplification, which will give an extra multiplicative overhead of order $\mathcal{O}(\kappa)$. 

It is of course unreasonable to expect that the eigenvalues of $A$ satisfy the above assumption.
Nonetheless, we expect that by taking $t_0$ sufficiently large, this assumption will approximately hold, and the algorithm will still work.
We now point out the main differences between the simple case considered above and the real HHL algorithm from \cite{Harrow2009QuantumAlgorithmLinearSystemsEquations}.
The clock of QPE is not prepared in a uniform superposition but in
\begin{equation}\label{eq:clock}
    \ket{\Omega}_{c} := \sqrt{\dfrac{2}{T}} \sum_{\tau = 0}^{T-1} \sin\Bigl({\dfrac{\pi (\tau+1/2)}{T}}\Bigr)\ket{\tau}_c\,.
\end{equation}
After applying QPE the state of the system is now 
\begin{equation}
    \qpe \bigl(\ket{\Omega}_c\otimes\ket{b}\bigr)=
    \sum_{j=1}^{N}\sum_{\tau=0}^{T-1}\beta_j\alpha_{\tau\vert j}\ket{\tau}\otimes \ket{u_j}\,,
\end{equation}
with 
\begin{equation}
    \alpha_{\tau\vert j}\coloneqq\frac{\sqrt{2}}{T}\sum_{\tau'=0}^{T-1}\sin\Bigl[\frac{\pi(\tau'+1/2)}{T}\Bigr]\exp\Bigl(i\frac{\tau'}{T}(\lambda_jt_0-2\pi\tau)\Bigr)\,.
\end{equation}
The amplitude $\alpha_{\tau\vert j}$ is picked around $\tau$ such that $\lambda_j\sim 2\pi/t_0 
\tau$, however it has nonzero tails for all other values of $\tau$.
The reason why the clock state is initialized in $\ket{\Omega}$ is that these tails are suppressed compared to those we would have using the uniform superposition. 
We summarize the algorithm in the following pseudocode. 
\begin{algorithm}
\caption{HHL}\label{alg:HHL}
\begin{algorithmic}[1]
\Function{HHL}{oracle $\mathcal{P}_A$, oracle $\mathcal{P}_b$, condition number $\kappa$, $\varepsilon >0$}
\State Prepare $\ket{b}$ with $\mathcal{P}_b$
\State Prepare the clock in $\ket{\Omega}$ from Eq.\ \eqref{eq:clock}
\State Apply $\mathcal{U}$ from Eq.\ \eqref{eq:hhl_U}
\State Apply inverse quantum Fourier transform
\State Append flag ancilla initialized in $\ket{0}$ and rotate it with $\mathcal{R}$ from Eq.\ \eqref{eq:hhl_R}
\State Undo steps (3), (4), and (5)
\EndFunction
\end{algorithmic}
\end{algorithm}

Note that we have not included the amplitude amplification step in the algorithm, because it is independent of the details of the algorithm. 
We will consider it separately further below.
However, it is important to provide an expression for the success probability of the flag measurement, i.e.\ a generalization of \eqref{eq:probFlagSimple} to the general case.
This is given by
\begin{equation}
    \tilde{p}=\sum_{j=1}^N\sum_{\tau=0}^{T-1}\abs{\beta_j}^2\abs{\alpha_{\tau\vert j}}^2\tilde{f}_\tau^2\,,
\end{equation}
where we have defined 
\begin{equation}
\tilde{f}_\tau \coloneqq 
    \begin{dcases}
        \frac{t_0}{4\pi \kappa}\frac{1}{\tau} \quad &\tau\ge \frac{t_0}{4\pi \kappa} \, ,\\
        0&\text{otherwise}\,.
    \end{dcases}
\end{equation}
To run the quantum algorithm, we don't need to know this probability, a lower bound is sufficient. 
However, to perform our hybrid benchmarking, and in particular to compute the overhead introduced by QAA in expectation, we need to compute this probability numerically.
However, the expression above for $\tilde{p}$ is rather annoying to compute, even numerically. Hence we provide the following $\varepsilon$-accurate estimate.
\begin{lemma}\label{lem:app_p_succ}
Let 
\begin{equation}\label{eq:hhl_p}
p\coloneqq\frac{1}{4\kappa^2}\norm{x}^2\,,
\end{equation}
we have that $\abs{p-\tilde{p}}<\varepsilon$, where $\varepsilon$ is the same accuracy used to run the HHL algorithm. 
\end{lemma}
Therefore, in our numerics, we can estimate the probability of success by only computing the condition number and the norm of the solution $x$.
This is a good approximation as long as $\varepsilon\ll 1/\kappa^2$. 

Using this approximation, we can also easily find a lower bound on the probability.
Noticing that $\norm{x}^2=\sum_j\abs{\beta_j}^2/\lambda_j^2$ and $\lambda_j\le 1$, we find that 
$p \ge (2\kappa)^{-2}$\,. Unsurprisingly, this is the lower bound we had already found in the simple example above, see Eq.\ \eqref{eq:low_bound_hhl_p_1}.

To complete the algorithm, we need to provide a value for $t_0$ such that the condition $t_0\lambda_j=2\pi \tau_j$ is approximately satisfied, and the post-selected state is $\varepsilon$-close to $\ket{x}$. 
\begin{lemma}\label{lem:hhl_t0}
The post-selected state is $\varepsilon$-close to $\ket{x}$ provided we pick
\begin{equation}\label{eq:hhl_t0}
    t_0 = c_F^{1/2} \frac{\kappa}{\varepsilon} \, ,
\end{equation}
where $c_F=2^7+22\pi^2+(64+14\pi^2)^2/\pi^2$.
\end{lemma}
The proof of this lemma is straightforward but somewhat long, see \autoref{sec:proof_lem_hhl}.

Having summarized the algorithm, we now proceed to estimate its cost. 
\LemmaCostHHL*
\begin{proof}
The only steps of the algorithm invoking $P_A$ 

are 
the steps in which we apply $\mathcal{U}$ or $\mathcal{U}^\dagger$. 
The query complexity of a general controlled unitary $\sum_j \lvert j \rangle \langle j\rvert \otimes U_j$ can be estimated by the query complexity of its most expensive subroutine $U_{j_0}$, which in this case is\footnote{Here, we are neglecting an irrelevant $(T-1)/T$ factor.} $e^{iAt_0}$, where $t_0$ is given in Eq.\ \ref{eq:hhl_t0}. 
As explained above, the algorithm is successful with a probability that is close to \eqref{eq:hhl_p} as long as $\varepsilon\ll 1/\kappa$.
We boost this probability to $\mathcal{O}(1)$ using QAA.
As explained in App.\ \ref{section:AppendixQAA}, this leads to an extra multiplicative overhead given by Eq.\ \eqref{n_q}. 
To compute this quantity we also need a lower bound for the probability of success, that as explained above can be taken to be $p_0=(2\kappa)^{-2}$.
\end{proof}

\subsection{Proof of Lemma \ref{lem:hhl_t0}}\label{sec:proof_lem_hhl}
The proof closely follows \cite{Harrow2009QuantumAlgorithmLinearSystemsEquations} (in particular, App.\ A of the supplementary material). The main difference is that we consider only well-conditioned matrices (set $g=0$ in their notation) and explicitly keep track of numerical constants. 

The final state of the system after post-selection and the inverse QPE is 
\begin{equation}
    \ket{\tilde{\psi}}=\frac{1}{\tilde{Z}}\qpe^\dagger \cdot \sum_{j=1}^N \beta_j \alpha_{\tau\vert j}\tilde{f}_\tau \ket{\tau}\otimes \ket{u_j} ,
\end{equation}
with normalization 
\begin{equation}
    \tilde{Z}^2=\sum_{j=1}^N\sum_{\tau=0}^{T-1} \abs{\beta_j}^2 \abs{\alpha_{\tau\vert j}}^2\tilde{f}_\tau^2\,,
\end{equation}
and recall that 
\begin{equation}
\tilde{f}_\tau = 
    \begin{dcases}
        \frac{t_0}{4\pi \kappa}\frac{1}{\tau} \quad &\tau\ge \frac{t_0}{4\pi \kappa} \, ,\\
        0&\text{otherwise}\,.
    \end{dcases}
\end{equation}
To simplify the notation, let $\tau_* \coloneqq t_0/(4\pi \kappa)$.
The correct output state would be $\ket{\psi}=\ket{0}_c\otimes \ket{x}$ with
\begin{equation}
    \ket{x}=\frac{1}{Z}\sum_{j=1}^N \beta_j f_j\ket{u_j},
\end{equation}
where we have defined $f_j\coloneqq(2k\lambda_j)^{-1}$, and normalization
\begin{equation}
    Z^2=\sum_{j=1}^N \abs{\beta_j}^2f_j^2\,.
\end{equation} 
For every $\lambda_j$ there is an integer $\tau_j$ and real $\delta_j$ with $\abs{\delta_j}\le 1/2$ such that $t_0\lambda_j=2\pi(\tau_j+\delta_j)$, so we can also write 
\begin{equation}
    f_j = \frac{t_0}{4\pi \kappa}\frac{1}{\tau_j+\delta_j}\,.
\end{equation}
We want to ensure that the overlap between these two states is large, a simple calculation shows that 
\begin{equation}
    F\coloneqq \braket{\psi|\tilde{\psi}}=\frac{1}{Z\tilde{Z}}\sum_{j=1}^N\sum_{\tau=0}^{T-1}\abs{\beta_j}^2\abs{\alpha_{\tau\vert j}}^2\tilde{f}_\tau f_j\,.
\end{equation}
Following \cite{Harrow2009QuantumAlgorithmLinearSystemsEquations}, we introduce $\llbracket * \rrbracket=\sum_{j=1}^N\sum_{\tau=0}^{T-1}\abs{\beta_j}^2\abs{\alpha_{\tau\vert j}}^2*\,$ and rewrite $F$ as
\begin{equation}
    F = \frac{\llbracket \tilde{f}f\rrbracket}{\sqrt{\llbracket \tilde{f}^2\rrbracket\llbracket f^2\rrbracket}}\,.
\end{equation}
Following the same steps as in (A12)-(A17) in \cite{Harrow2009QuantumAlgorithmLinearSystemsEquations}, one can show that 
\begin{equation}\label{eq:bound_F_1}
    F \ge 1- \frac{\llbracket (\tilde{f}-f)^2\rrbracket}{2\llbracket f^2\rrbracket}
    \biggl(1+ \frac{\llbracket (\tilde{f}-f)f\rrbracket}{2\llbracket f^2\rrbracket}\biggr)-\biggl( \frac{\llbracket (\tilde{f}-f)f\rrbracket}{2\llbracket f^2\rrbracket}\biggr)^2\,.
\end{equation}
We need to bound $\llbracket (\tilde{f}-f)^2\rrbracket$ and $\llbracket (\tilde{f}-f)f\rrbracket$. 

To do this, we use that 
\begin{equation}\label{eq:bound_diff_fs}
   (\tilde{f}-f)^2\le \frac{4\kappa^2}{t_0^2}\Delta^2f^2\,,
\end{equation}
where 
\begin{equation}
\begin{split}
    \Delta &\coloneqq t_0\lambda_j -2\pi\tau\\
    &=2\pi(\tau_j-\tau+\delta_j)\,.
\end{split}
\end{equation}
This can be proven as follows. First consider $\tau\ge \tau_*$, then
\begin{equation}
\begin{split}
    (\tilde{f}-f)^2 &=\frac{\Delta^2}{16\pi^2\kappa^2}\frac{1}{\tau^2\lambda_j^2}\\
    &\le \frac{1}{t_0^2}\frac{\Delta^2}{\lambda_j^2} \, ,
\end{split}
\end{equation}
where we have used the definition of $f$, $\tilde{f}$, and $\Delta$, in the first line.
In the second line, we used $\tau\ge \tau_*$. 
Finally, plugging the decomposition of $\lambda_j$ in $\tau_j$ and $\delta_j$, we arrive at \eqref{eq:bound_diff_fs}.
For $\tau< \tau_*$, we simply have $(\tilde{f}-f)^2=f^2$. However, we also have $\lambda_j-2\pi/t_0\tau\ge 1/2\kappa$ (which follows from $\lambda_j\ge 1/\kappa$), so $\Delta\ge t_0/2\kappa$ and \eqref{eq:bound_diff_fs} still apply.

Let us begin by considering $\llbracket (\tilde{f}-f)^2\rrbracket$. Using \eqref{eq:bound_diff_fs}, we can bound it with $\llbracket f^2\Delta^2\rrbracket$ times a constant. 
We have
\begin{equation}
    \llbracket f^2\Delta^2\rrbracket=4\pi^2\sum_{j=1}^N\abs{\beta_j}^2f^2\sum_{\tau=0}^{T-1}\abs{\alpha_{\tau\vert j}}^2(\tau_j-\tau+\delta_j)^2\,
\end{equation}
We split the sum over $\tau$ in three pieces, $\tau<\tau_j-1$, $\tau\in \{\tau_j-1, \tau_j, \tau_j+1\}$ and $\tau>\tau_j+1$. In other words, 
\begin{equation}
    \sum_{\tau=0}^{T-1}\abs{\alpha_{\tau\vert j}}^2(\tau_j-\tau+\delta_j)^2 = S_<+S_j+S_>\,,
\end{equation}
with 
\begin{align}
    S_< \coloneqq &\sum_{\tau=0}^{\tau_j-2}\abs{\alpha_{\tau\vert j}}^2(\tau_j-\tau+\delta_j)^2\,,\\
    S_j \coloneqq & \abs{\alpha_{\tau_j-1\vert j}}^2(1+\delta_j)^2 + \abs{\alpha_{\tau_j\vert j}}^2\delta_j^2 + \abs{\alpha_{\tau_j+1\vert j}}^2(1-\delta_j)^2\,,\\
    S_> \coloneqq &\sum_{\tau=\tau_j+2}^{T-1}\abs{\alpha_{\tau\vert j}}^2(\tau_j-\tau+\delta_j)^2\,.
\end{align}
Using $\abs{\alpha_{\tau\vert j}}^2\le 1$ and $\abs{\delta_j}\le 1/2$, we can bound $S_j\le 11/4$. 
For the other two terms, we use that 
\begin{equation}
    \abs{\alpha_{\tau\vert j}}^2\le \frac{4}{\pi^2}\biggl(\frac{t_0}{2\pi}\lambda_j-\tau\biggr)^{-4}\,,
\end{equation}
which is proven in App.\ A3 in \cite{Harrow2009QuantumAlgorithmLinearSystemsEquations}.
For example, we have 
\begin{equation}
\begin{split}
    S_> &\le \frac{4}{\pi^2}\int_{\tau_j+1}^{\infty}\frac{1}{(\tau_j-\tau+\delta_j)^2} \; \mathrm{d}\tau \\
    &=\frac{4}{\pi^2}\frac{1}{1-\delta_j}\le \frac{8}{\pi^2}\,,
\end{split}
\end{equation}
where in the first step, we have bounded the sum by an integral, and in the last we have used that $\abs{\delta_j}\le 1/2$. 
Similarly, it can be shown that $S_<\le 8/\pi^2$. 
Plugging this into the expression for $\llbracket f^2\Delta^2\rrbracket$ we find that 
\begin{equation}
   \llbracket f^2\Delta^2\rrbracket\le (2^6+11\pi^2)\llbracket f^2\rrbracket\,, 
\end{equation}
which leads to
\begin{equation}
    \frac{\llbracket (\tilde{f}-f)^2\rrbracket}{2\llbracket f^2\rrbracket}\le (2^7+22\pi^2)\frac{\kappa^2}{t_0^2}\,.
\end{equation}

Next, we consider $\llbracket (\tilde{f}-f)f\rrbracket$. We have
\begin{equation}
\begin{split}
    \llbracket (\tilde{f}-f)f\rrbracket &\le \Bigl\llbracket \sqrt{(\tilde{f}-f)^2f^2}\Bigr\rrbracket\\
    &\le \frac{2\kappa}{t_0}\llbracket \abs{\Delta}f^2\rrbracket\,,
\end{split}
\end{equation}
where in the second line we have used \eqref{eq:bound_diff_fs}. 
To bound $\llbracket \abs{\Delta}f^2\rrbracket$, we follow the same steps as for $\llbracket \Delta^2f^2\rrbracket$. The result is 
\begin{equation}
  \llbracket \abs{\Delta}f^2\rrbracket  \le \Bigl( \frac{32}{\pi}+7\pi\Bigr)\llbracket f^2\rrbracket\,.
\end{equation}

Plugging everything in \eqref{eq:bound_F_1}, we arrive at 
\begin{equation}
    F \ge 1- c_F \frac{\kappa^2}{t_0^2} + O\Bigl(\frac{\kappa^3}{t_0^3}\Bigr)\,.
\end{equation}
From this it is clear that $t_0=\mathcal{O}(\kappa/\varepsilon)$, and the last term gives a negligible contribution to $F$, if $\varepsilon$ is small. Hence, we will neglect it. Finally, setting the leading order correction equal to $\varepsilon^2$, we arrive at the claimed result.

\subsection{Proof of Lemma \ref{lem:app_p_succ}}
The proof is very similar to the proof of Lemma \ref{lem:hhl_t0}. First, notice that 
\begin{equation}
    p=\sum_{j=1}^N\abs{\beta_j}^2f_j^2=\llbracket f^2\rrbracket\,,
\end{equation}
where $f_j\coloneqq(2k\lambda_j)^{-1}$, and $\llbracket\cdot\rrbracket$ are defined in the proof of Lemma \ref{lem:hhl_t0}. 
Similarly, we have that $\tilde{p}=\llbracket \tilde{f}^2\rrbracket$. 
In this notation, what we would like to bound is $\lvert\llbracket \tilde{f}^2\rrbracket-\llbracket f^2\rrbracket\rvert$.
A little algebra shows that 
\begin{equation}
    \lvert\llbracket \tilde{f}^2\rrbracket-\llbracket f^2\rrbracket\rvert\le \llbracket (\tilde{f}-f)^2\rrbracket+2\llbracket\lvert\tilde{f}-f\rvert f\rrbracket\,.
\end{equation}
These quantities were already bounded in the proof of Lemma \ref{lem:hhl_t0}, namely
\begin{equation}
\begin{split}
\llbracket (\tilde{f}-f)^2\rrbracket&\le \frac{k^2}{t_0^2}2(2^7+22\pi^2)\llbracket f^2\rrbracket\,,\\
\llbracket\lvert\tilde{f}-f\rvert f\rrbracket &\le \frac{k}{t_0}2\Bigl(\frac{32}{\pi}+7\pi\Bigr)\llbracket f^2\rrbracket\,.
\end{split}
\end{equation}
Since $k/t_0=\varepsilon/c_F^{1/2}$ and $\llbracket f^2\rrbracket\le 1/4$, which can be shown using $\lambda_j\ge 1/k$, we have that 
\begin{equation}
\begin{split}
    \lvert\llbracket \tilde{f}^2\rrbracket-\llbracket f^2\rrbracket\rvert&\le \frac{32/\pi+7\pi}{c_F^{1/2}}\varepsilon+\mathcal{O}(\varepsilon^2)\\
    &\le \frac{\varepsilon}{2}+\mathcal{O}(\varepsilon^2)\\
    &\le \varepsilon \, ,
\end{split}
\end{equation}
where we have used the expression for $c_{F}^{1/2}$ in the second line.
In the last line, we have assumed that $\varepsilon$ is sufficiently small such that the $\mathcal{O}(\varepsilon^2)$ term is smaller than $\varepsilon/2$.

\section{\label{section:AppendixQLSFourier}QLS-Fourier}
By QLS-Fourier we denote the algorithm from Theorem 3 in \cite{Childs2017QuantumAlgorithmSistemyLinearEquations}. 
Here, we briefly summarize how the algorithm works and give a formula for the query complexity of the algorithm including constant prefactors.

The idea of QLS-Fourier is to approximate the matrix $A^{-1}$ by a linear combination of unitaries.
Namely the authors of \cite{Childs2017QuantumAlgorithmSistemyLinearEquations} show that
\begin{equation}
    h(A) = \sum_{j=0}^{J-1}\sum_{l=-L}^{L}\alpha_{jl}\cdot U_{jl}
\end{equation} 
with
\begin{equation}\label{eq:lcu_fourier}
\begin{split}
    \alpha_{jl}&=\frac{1}{\sqrt{2\pi}}\Delta_y\Delta_z^2\abs{l}e^{-(l\Delta_z)^2/2},\\
    U_{jl}&=\sign(l)\cdot e^{-iAt_{jl}} \, ,\quad t_{jl} = jl\Delta_y\Delta_z
\end{split}
\end{equation}
satisfies $\norm{A^{-1}-h(A)}\le \varepsilon$ ($\norm{\cdot}$ being the operator norm) provided one chooses $\Delta_y$, $\Delta_z$, $J$ and $L$ with a suitable asymptotic scaling.
According to these restrictions, we choose
\begin{equation}
\begin{split}
    \Delta_y &=\frac{\varepsilon}{16}\Bigl[\log\Bigl(1+\frac{8\kappa}{\varepsilon}\Bigr)\Bigr]^{-1/2}\,,\\ 
    J&=\Bigl\lfloor\frac{16\sqrt{2}\kappa}{\varepsilon}\log\Bigl(1+\frac{8\kappa}{\varepsilon}\Bigr)\Bigr\rfloor\,,\\
    \Delta_z &=\frac{2\pi}{\kappa+1}\Bigl[\log\Bigl(1+\frac{8\kappa}{\varepsilon}\Bigr)\Bigr]^{-1/2}\,,\\ 
    L&=\Bigl\lfloor\frac{\kappa+1}{\pi}\log\Bigl(1+\frac{8\kappa}{\varepsilon}\Bigr)\Bigr\rfloor \, .
\end{split}
\end{equation}
For a proof of this statement, we refer the reader to the original article \cite{Childs2017QuantumAlgorithmSistemyLinearEquations}, and for more details on our choice to the review provided in App.\ C of \cite{Ammann2023RealisticRuntimeAnalysisQuantumSimplex}.

Linear combination of unitaries (LCU)
is briefly summarized in App.\ \ref{sec:LCU}, a formula for its query complexity is given in Lemma \ref{lemma:lcu}. 
In this case, the unitaries to implement are of the form $e^{iAt}$, for various values of $t > 0$.
They can be implemented using Hamiltonian simulation algorithms.
In this section, we leave the Hamiltonian simulation algorithm unspecified.
To generate our benchmarks, we have used the algorithm for quantum simulation of \cite{Low2019Qubitization}, which is reviewed in App.\ \ref{section:AppendixHamiltonianSimulation}.
Finally, one has to boost the success probability of LCU to $\mathcal{O}(1)$ by quantum amplitude amplification, which is reviewed in App.\ \ref{section:AppendixQAA}. 
To implement QAA, we need a lower bound $p_0$ on the success probability.
As showed in \cite{Childs2017QuantumAlgorithmSistemyLinearEquations} (see also the proof of Lemma \ref{lemma:CostQLS-Fourier} below), we can use
\begin{equation}
    p_0 = \frac{1}{\alpha^2}\,,
\end{equation}
where $\alpha = \sum_{jl}\alpha_{jl}$.  
Computing the sum over $j$ in the definition of $\alpha$, we find that 
\begin{equation}\label{eq:alpha}
    \alpha=4\sqrt{\pi}\frac{\kappa}{\kappa+1}\sum_{l=1}^{L}l\Delta_z e^{-(l\Delta_z)^2/2}\,,
\end{equation} 
which can be easily computed numerically if we know $\kappa$.
We summarize the steps of the algorithm in Alg.\ \ref{alg:qlsa_four}.
\begin{algorithm}
\caption{QLS-Fourier}\label{alg:qlsa_four}
\begin{algorithmic}[1]
\Function{QLS-Fourier}{matrix $A$, vector $b$, condition number $\kappa$, $\varepsilon >0$}
\State Prepare $\ket{b}$ with $\mathcal{P}_b$
\State Apply LCU$(\{U_{jl}\}, \{\alpha_{jl}\}, J(2L+1))$

\Comment{$U_{jl}$, $\alpha_{jl}$ from Eq.\ \eqref{eq:lcu_fourier}}
\State Apply QAA(LCU, $\alpha$, $\log_2(J(2L+1))$)
\EndFunction
\end{algorithmic}
\end{algorithm}

Finally, the query complexity of the algorithm can be computed using the following lemma. 

\LemmaCostQLSF*
\begin{proof}
The only term querying $\mathcal{P}_A$ is LCU. 
Considering the steps needed to implement the LCU, as explained in App. \ref{sec:LCU}, we see that the only step in LCU that queries $\mathcal{P}_A$ is the controlled unitary
\begin{equation}
    U = \sum_{jl}\ketbra{jl}\otimes e^{-iAt_{jl}}\,.
\end{equation}
The query complexity of such a controlled unitary can be estimated by the query complexity of its most expensive term, that in this case is
\begin{equation}
    U_{JL}=e^{-iA JL\Delta_y\Delta_z}.
\end{equation}
In other words, the query complexity of QLS-Fourier is given by the query complexity of Hamiltonian simulation, with Hamiltonian $A$, for time 
\begin{equation}
\begin{split}
    t &= JL\Delta_y\Delta_z\\
    & = 2\sqrt{2}\kappa \log\Bigl(1+\frac{8\kappa}{\varepsilon}\Bigr).
\end{split}
\end{equation} 
Finally, LCU is successful only with probability
\begin{equation}
    p = \frac{\norm{h(A)\ket{b}}^2}{\alpha^2}\,,
\end{equation}
where $\alpha = \sum_{jl}\alpha_{jl}$. 
Approximating the sum in Eq. \eqref{eq:alpha} by an integral \cite{Childs2017QuantumAlgorithmSistemyLinearEquations}, one can see that $\alpha=\mathcal{O}(\kappa\sqrt{\log(\kappa/\varepsilon)})$.
We can boost this probability to $\mathcal{O}(1)$ using QAA, provided we know in advance a lower bound on this success probability. 
To see this, first recall that the parameters of the algorithm were chosen such that $\norm{h(A)-A^{-1}}\le \varepsilon$, so up to error $\mathcal{O}(\varepsilon)$ we have 
\begin{equation}\label{eq:p_four_app}
    p = \frac{\norm{A^{-1}\ket{b}}^2}{\alpha^2}\,.
\end{equation}
Then we can use that $\norm{A^{-1}\ket{b}}\ge 1$, which follows from $\norm{A}=1$. 
So we can lower bound the probability of success as $p\ge p_0$ with 
\begin{equation}
    p_0 = \frac{1}{\alpha^2} \, .
\end{equation}
As explained in App. \ref{section:AppendixQAA}, QAA  introduces a constant overhead to the query complexity which can be computed (in expectation value), provided that we can exactly compute the probability of success.
From the discussion above, we have that up to $\mathcal{O}(\varepsilon)$ corrections, we can compute the probability of success as 
\begin{equation}
    p=\frac{\norm{x}^2}{\alpha^2}\,,
\end{equation}
where we have assumed that $b$ is normalized and $x=A^{-1}b$.
\end{proof}

\section{\label{section:AppendixQLSChebyshev}QLS-Chebyshev}
\begin{algorithm}
\caption{QSL-Chebyshev from \cite{Childs2017QuantumAlgorithmSistemyLinearEquations}}\label{alg:qlsa_cheby}
    \begin{algorithmic}[1]
        \Function{QLS}{state  $\ket{b}$, sparse matrix $A$ with sparsity $d$, $\varepsilon >0$, condition number bounded by $\kappa$}
        \State Run $\text{LCU}(\ket{b}, \{\alpha_j\}, \{U_{j}\}) \eqqcolon \ket{\psi}$\Comment{$\alpha_j=4\frac{\sum_{i=j+1}^b{2b \choose b+i}}{2^{2b}}$, where $b=(d\kappa)^2\log_2(d\kappa/\varepsilon)$}
        \State Amplify($\ket{\psi}$) with QAA\Comment{$\{U_{i}\}$ is defined in \eqref{eq:Ui_def_cheby}}
        \State Return $\ket{\psi_i}$, obeying $\left|\left|\frac{A^{-1}\ket{b}}{\left|\left|A^{-1}\ket{b}\right|\right|}-\ket{\psi_i}\right|\right|\leq \varepsilon$
        \EndFunction
    \end{algorithmic}
\end{algorithm}

The Chebyshev approach relies on two observations.
Firstly, we can approximate $1/x$ on the relevant domain $\mathcal{D}_{d\kappa}=[-1, \frac{-1}{d\kappa})\cup (\frac{1}{d\kappa}, 1]$ with a linear combination of Chebyshev polynomials of the first kind.
Secondly, we can enact a linear combination of such Chebyshev polynomials with a quantum walk.
More specifically,
\begin{equation}
    g(x):=4\sum_{j=0}^{j_0}(-1)^j\left[\frac{\sum_{i=j+1}^b{2b \choose b+i}}{2^{2b}}\mathcal{T}_{2j+1}(x)\right]
    \label{eq:chebyshev_fcn}
\end{equation}
is $2\varepsilon$ close to $1/x$ on $\mathcal{D}_{d\kappa}$, given $b=\lceil(d\kappa)^2\log_2(d\kappa/\varepsilon)\rceil$ and $j_0=\lceil\sqrt{b\log_2(4b/\varepsilon)}\rceil$.
To encode $g(x)$ in Alg. \ref{alg:lcu}, we require $\alpha_i>0$.
This is not a problem, since $e^{i \varphi} U$ is unitary for any unitary $U$ and $\varphi \in \R$.
Since we actually need to work with $g(x)/d$, we define
\begin{align}
    \alpha_j&=\frac{4}{d}\frac{\sum_{i=j+1}^b{2b \choose b+i}}{2^{2b}} \, .\nonumber\\
\end{align}
There are $j_0+1$ terms, up to degree $2j_0+1$. 
For $x\in \mathcal{D}_{d\kappa}$, we have $\left|g(x)\right|\geq 1$, $\left|g(A)-\sum_i\alpha_i\mathcal{T}_i(A)\right|\leq \varepsilon\in(0, 1/2)$. 
The detailed derivation of this result is given in \cite{Childs2017QuantumAlgorithmSistemyLinearEquations}, and we refer readers there for details. 
Given a state preparation oracle $\mathcal{P}_b$ building $\ket{b}$, then $A^{-1}\ket{b}$ can be approximated by $g(A/d)$, and specifically
\begin{equation}
    \left|\left|\frac{A^{-1}\ket{b}}{\left|\left|A^{-1}\ket{b}\right|\right|}-\frac{g(\frac{A}{d})\ket{b}}{\left|\left|g(\frac{A}{d})\ket{b}\right|\right|}\right|\right|\leq 4\varepsilon \, .
\end{equation}
We will be able to implement $g(x)$  using LCU - specifically, we will embed the operator set $\{\mathcal{T}_i\}$ within a set of unitary matrices, and then the LCU will build an $\varepsilon$-close approximation to $g(x)$.

The Chebyshev terms in the above approximation can be constructed with quantum walk operators for the Hamiltonian $A'=A/d$ in the space $\mathbb{C}^{2N}\otimes \mathbb{C}^{2N}$, provided that $A$ is Hermitian with eigenvalues $\lambda\in \mathcal{D}_{\kappa}=\left[-1, -\kappa^{-1}\right)\cup \left(\kappa^{-1}, 1\right]$. 
Note that this implies that $A'$ has eigenvalues in the range $\lambda\in \mathcal{D}_{d\kappa}=\left[-\frac{1}{d}, -\frac{1}{d\kappa}\right)\cup \left(\frac{1}{d\kappa}, \frac{1}{d}\right]$, so we need to use an approximation valid in the range $\mathcal{D}_{d\kappa}$. 
We also require the largest entry of $A$ in absolute value obeys $\left|\left|A\right|\right|_{\max}\leq 1$ where the $N\times N$ matrix $A$ is $d$-sparse.\footnote{We can get around this by defining $A'=A/d||A||_{\text{max}}$. This is built into QSVT matrix inversion algorithms.} Use a set of states encoding the elements of $A$
\begin{equation}
\begin{split}
     \ket{\psi_j}
     \coloneqq
     \ket{j}\otimes \frac{1}{\sqrt{d}}
     \cdot \sum_{k\in\left[N\right]:A_{jk}\neq 0}
     \biggl(&\sqrt{A_{jk}^*}\ket{k}  \\ 
     +&\sqrt{1-\left|A_{jk}\right|}\ket{k+N} \biggr) \, .   
\end{split}
\end{equation}
Note that a consistent convention choice for the square root is required when $A_{jk}$ is complex \cite{Berry2015HamiltonianSimulationNearlyOptimalDependenceOnAllParameters}. Following \cite{Childs2017QuantumAlgorithmSistemyLinearEquations}, we assume that the oracle returns exactly $d$ nonzero entries for any $j$.\footnote{If row $j$ has less than d non-zero elements, we assume the oracle can return some $d$ elements $A_{jk}=0$.} 
We construct an isometry $T:\mathbb{C}^{N}\rightarrow \mathbb{C}^{2N}\otimes \mathbb{C}^{2N}$, and swap and walk operators $S, W: \mathbb{C}^{2N}\otimes \mathbb{C}^{2N} \rightarrow 
\mathbb{C}^{2N}\otimes \mathbb{C}^{2N}$:
\begin{align}
    T &\coloneqq \sum_{j'\in\left[N\right]}\ket{\psi_{j'}}\bra{j'} ,\\
    S &\coloneqq S\ket{j, k}=\ket{k, j} ,\\
    W &\coloneqq S\left(2TT^{\dag}-{I}\right).
\end{align}   
With initial state $\ket{\psi}$, the action of $W^{i
}$ on $T\ket{\psi}$ is
\begin{equation}
W^iT\ket{\psi}=\mathcal{T}_i(A')T\ket{\psi}+\ket{\perp_{\psi}}.
\end{equation}
$\ket{\perp_{\psi}}$ is a (here unnormalized) state orthogonal to $T\ket{j}$ for all $j\in[N]$. 
See Lemma 16 in \cite{Childs2017QuantumAlgorithmSistemyLinearEquations} for the proof.

To implement $T$ as a unitary, we introduce $t=\log_2(\lceil 2N\rceil)+1$ ancillary qubits and define $T_U: \mathbb{C}^{2N}\otimes\mathbb{C}^{2N}$ as the unitary operation  such that $\ket{0^t}\ket{\psi}\rightarrow T\ket{\psi}$.

We can now use Alg. \ref{alg:lcu} to build $g(A')$ as a quantum circuit, and by doing so obtain
\[
\left|\left|dA^{-1}-g(A')\right|\right|\leq {\varepsilon} \quad \Rightarrow \quad \left|\left|A^{-1}-\frac{1}{d}g(A')\right|\right|\leq \frac{\varepsilon}{d}\leq \varepsilon \, .
\]
So below, we will assume the function we want prepared is $\frac{g(A')}{d}$, an $\varepsilon$-close approximation of $A^{-1}$. 
Doing so with a LCU (see Section \ref{sec:LCU}), we will obtain an $4\varepsilon$-approximation to the desired state $A^{-1}\ket{b}$. 
To use Alg. \ref{alg:lcu}, we define 
\begin{equation}
    U_{j}=(-1)^jT_U^{\dag}W^{2j+1}T_U 
\end{equation}
so that
\begin{equation}
    U_j\ket{0^{\otimes t }}\ket{\psi}=e^{-ij\pi}\ket{0^{\otimes t}}\mathcal{T}_{2j+1}(A)\ket{\psi}+\ket{\Phi_{i}^{\perp}}.\label{eq:Ui_def_cheby}
\end{equation}
Then,
\begin{align}
V\ket{0^{m}}&=\frac{1}{\sqrt{\alpha}}\sum_{i=0}^{ j_0}\sqrt{\alpha_i}\ket{i},\\
 U&=\sum_{i}\ketbra{i}\otimes U_i \, ,\\
    \alpha&=\sum_i\alpha_i\leq \frac{4j_0}{d}\label{eq:cheby_alpha_bound} \, .
\end{align}
\eqref{eq:cheby_alpha_bound} follows from bounding $\alpha_i$'s in \eqref{eq:chebyshev_fcn}, see Lemma 19 in \cite{Childs2017QuantumAlgorithmSistemyLinearEquations} for more details. 
From inspection, $V^{\dag}UV$ enacts $\frac{1}{\alpha}\sum_i\alpha_iU_i$, and thus encodes $\frac{1}{\alpha}\sum_i\alpha_i\mathcal{T}_i$ within the circuit. 
Altogether,
\begin{equation}
\begin{aligned}
    V^{\dag}UV\ket{0^m}\ket{0^{t}}T\ket{b}&=\frac{1}{\alpha}\ket{0^{m+t}}\sum_ie^{-ij\pi}\alpha_i\mathcal{T}_{2i+1}(A)T\ket{b}\\
    &+\ket{\Phi^{\perp}}.  
\end{aligned}
    \label{QLSA_cheby_U}
\end{equation}

\subsection{Step costs}
$V$ is interpreted as a state preparation map \cite{Childs2017QuantumAlgorithmSistemyLinearEquations} on $m=\lceil\log_2 (j_0+1)\rceil$ qubits. 
To implement it, we use the algorithm of \cite{Mottonen2004TransformationQuantumStateUniformlyControlledRotations} which gives a procedure for computing this with $4(j_0-\log_2(j_0+1))$ CNOT gates plus a similar number of 1-qubit gates. 
The cost of $\cost[V]$ is still less than $\cost[U]$, and so the cost of a QLS step will not be dominated by the $V$ terms. Thus, we drop it to obatin a lower bound.
\subsubsection{Cost of W}
The costs of $W$ are given in Lemma 10 of \cite{Berry2015HamiltonianSimulationNearlyOptimalDependenceOnAllParameters}, but with notation $\mathcal{P}_A$ which as in \cite{Berry2015HamiltonianSimulationNearlyOptimalDependenceOnAllParameters} represents a call to either $\mathcal{O}_F$ or $\mathcal{O}_A$ defined in Section \ref{sec:block_encodings}. Preparing $T$ can be done with $\lceil \log_2 d\rceil $ (instead of $\log_2 N$) Hadamard gates, and then one call to $\mathcal{P}_A$.

Finally, $S$ can be implemented by individually swapping the $i^{th}$ qubit of register $j$ with the $i^{th}$ qubit of register $k$, so the cost is $\log_2(2N)\cost_2$. 
Then $T, W$ have costs 
\begin{align}
\cost{[T]}&=\cost[\mathcal{P}_A]+\lceil\log_2 d\rceil\cost_1\\
\cost{[W]}&=2\cost[\mathcal{P}_A]+2\log_2 d\cost_1+\log_2(2N)\cost_2 \, ,
\label{cost_W_Cheby}
\end{align}
\subsubsection{Cost of U}
We use Lemma 8 in \cite{Childs2017QuantumAlgorithmSistemyLinearEquations}, and will take the cost of implementing any $U_i$ as a lower bound on the cost of implementing controlled operation $CU_i$. 
We can write $U$ as
\begin{equation}
    U=\left(I\otimes T_U^{\dag}\right)W\left(\sum_{j=0}^{j_0}\ketbra{j}\otimes (-1)^j\left(W^{2}\right)^j\right)\left(I\otimes T_U\right)
\end{equation}
In the oracle model, we do not assume any special structure that would justify $\cost{[W]}=\cost{[W^{2^j}]}$, or any other reduced cost. 
Hence, $\cost{[W^{2^j}]}=2^j\cost{[W]}$.
Ignoring some 1-qubit gates to set the coefficient signs\footnote{To get the correct signs on each $U_i$, we would apply 1-qubit gates $e^{i\phi_j\sigma_z}$ operations to half of the $CW$ steps. 
However, this is not a leading cost, so we drop it.}, the key point is to realize that we can implement this as a sum of controlled operations, controlled on the first register. 
Then, we use the following lemma.
\begin{lemma}[Corollary 8 from \cite{Childs2017QuantumAlgorithmSistemyLinearEquations}]
    Let $U=\sum_{i=0}^N\ketbra{i}\otimes Y^i$, where $Y$ is a unitary with gate complexity $G$. Let the gate complexity of $Y^{2^j}$ be $G_j\leq 2^{j}G$. Then the gate complexity of $U$ is $\mathcal{O}\left(\sum_{j=0}^{\lfloor\log N\rfloor}G_j\right)=\mathcal{O}(NG)$.
\end{lemma}
\begin{proof}
    We consider unitaries $Y^{2^j}$, $j\in\{0, 1,...\lfloor\log(N)\rfloor\}$, with gate complexity $G_j$. Unlike \cite{Childs2017QuantumAlgorithmSistemyLinearEquations}, we will assume $G_j=2^{j}G$. 
    Like \cite{Childs2017QuantumAlgorithmSistemyLinearEquations}, we will work with $G_j$ as a lower bound on the cost of $Y^{2^j}$. 
    To implement $U$, we perform controlled $Y^{2^j}$ operations: the first register is the control and $Y^{2^j}$ is applied on the second register. 
    The gate complexity is 
    \[
    \cost{[U]}\geq\sum_{j=0}^{\lfloor\log N\rfloor}G_j=\sum_{j=0}^{\lfloor\log N\rfloor}2^jG=\left(2^{\lfloor\log N\rfloor +1}-1\right)G  \, .
    \]
\end{proof}
With this result it is easy to see that
\begin{align}\label{U_costs_cheby}
    \cost{[U]} &\geq 2\cost{[T_U]}+\cost{[W]}+\left(2^{\lfloor\log j_0\rfloor +1}-1\right)\cost{[W^2]}\nonumber\\
    &=2\cost{[T_U]}+\cost{[W]}+2\left(2j_0 -1\right)\cost{[W]}\nonumber\\
    &=2\cost{[T_U]}+\left(4 j_0 -1\right)\cost{[W]}  \, .
\end{align}

\subsubsection{Costs of QLS-Chebyshev step}
\LemmaCostQLSC*
\begin{proof}
    Combining \eqref{U_costs_cheby} and \eqref{cost_W_Cheby} for $U_{\qlsC}=V^{\dag}UV\mathcal{P}_b \, $, 
\begin{align}
    \cost{[U_{\qlsC}]}
    &= 2\left(4j_0\right)\cost{[\mathcal{P}_A]} \, .
\end{align}
Then from the definition of our function $g(x)$,
\begin{equation}
    j_0=\left\lceil \sqrt{\left\lceil d\kappa\log_2\left(d\kappa/\varepsilon\right)\right\rceil \log_2\left[\frac{4}{\epsilon}\left\lceil d^2\kappa^2\log_2\left(d\kappa/\varepsilon\right)\right\rceil \right]}\right\rceil \, .
\end{equation}
A measurement of $\ket{0^{\lceil\log_2 N\rceil+1}}$ on the ancillary registers for LCU and the quantum walk, respectively with $m=\lceil\log_2 (j_0+1) \rceil, t=\lceil\log_2 (N) \rceil + 2$ qubits, prepares normalized state $\frac{\sum_i\alpha_i\mathcal{T}_i(A)T\ket{b}}{\left|\left|\sum_i\alpha_i\mathcal{T}_i(A)T\ket{b}\right|\right|}$. 
This  will occur with probability lower-bound by $\frac{1}{\alpha^2}\left|\left|\sum_i\alpha_i\mathcal{T}_i(A)T\ket{b}\right|\right|^2$. 
We must finally use amplitude amplification to increase the probability to $\mathcal{O}(1)$.
We will use quantum amplitude amplification as defined in \autoref{section:AppendixQAA}. 
\end{proof}

\section{\label{sec:block_encodings}Block Encodings}
\subsection{Quantum Walk Operators}
We review  the quantum walk techniques introduced in \cite{Childs_2009}, and applied in slightly modified settings in  \cite{BerryChilds2012BlackBoxSimulation, Berry2015HamiltonianSimulationNearlyOptimalDependenceOnAllParameters}. 
The quantum walk oracle is $W=iS\left(2TT^{\dag}-I\right)$, specified by isometry $T$. 
$T$ is defined by the state preparation procedure for $\ket{\phi_j}$  in Lemma \ref{lemma:encodingdsparsehamiltonians}, which relies on the sparse oracles equation \eqref{equation:SparseOracles}. This quantum walk oracle is used as an oracle in QLS-Chebyshev - see \autoref{section:AppendixQLSChebyshev} for a more comprehensive discussion of how to construct $T$ and $W$. $W$'s eigenvalues have the form $\pm e^{\pm i\arcsin\lambda}$. Note that $W$, which is constructed with $6$ calls to sparse-access oracles, could also be used in the Hamiltonian simulation procedure introduced in \cite{Low2017OptimalHamiltonianSimulationQuantumSignalProcessing} and discussed in \autoref{section:AppendixHamiltonianSimulation}.

\begin{lemma}[Encoding of $d$-sparse Hamiltonians)
    \cite{BerryChilds2012BlackBoxSimulation}]
    The $n$-dimensional quantum state 
    \begin{equation}
        \left|\phi_j\right\rangle = \frac{1}{\sqrt{d}} \sum_{k=1}^{n}\ket{k}\left[\sqrt{\frac{A_{j k}^*}{X}}|0\rangle+\sqrt{1-\frac{\left|A_{j k}\right|}{X}}|1\rangle\right],
    \end{equation}
    where $X=\frac{\Lambda_1}{ \varepsilon d}$ can be prepared using one call to the oracle $\mathcal{O}_F$ and two calls to the oracle $\mathcal{O}_A$ provided $\varepsilon \in$ $\left(0, \frac{\Lambda_1}{d \Lambda_{\max }}\right]$, with $\Lambda_{\max} \geq\|A\|_{\max }$ and $\Lambda_1 \geq\|A\|_1$ are known upper bounds on $\|A\|_{\max}$ and $\|A\|_1$.
    \label{lemma:encodingdsparsehamiltonians}
\end{lemma}
We now summarize the proof of Lemma \ref{lemma:encodingdsparsehamiltonians}. We use the $d$-sparse Hamiltonian encoding given in \cite{BerryChilds2012BlackBoxSimulation}. 
The steps of the state preparation protocol are given as algorithm \ref{alg:phij}.
\begin{algorithm}
\caption{Prepare $\ket{\phi_j}$}\label{alg:phij}
\begin{algorithmic}[1]
\Function{lemma10}{oracles $\mathcal{O}_F, \mathcal{O}_A$, sparsity $d$, index label $j$}
\State Initialize one ancilla qubit to $|0\rangle$
\State Prepare an equal superposition over $|1\rangle$ to $|d\rangle$ in the first register  $\frac{1}{\sqrt{d}} \sum_{k=1}^d\ket{k}|0\rangle$
\State Query the black box $\mathcal{O}_A$ to obtain $
\left|\phi_j^a\right\rangle:=\frac{1}{\sqrt{d}} \sum_{k \in F_j}\ket{k}|0\rangle$, where $F_j$ is the set of indices given by $\mathcal{O}_F$ on input $j$
\State  Query the black box oracle $\mathcal{O}_A$ to transform $\left|\phi_j^a\right\rangle$ to $
\frac{1}{\sqrt{d}} \sum_{k \in F_j}\ket{k}\left[\sqrt{\frac{A_{j k}^*}{X}}|0\rangle+\sqrt{1-\frac{\left|A_{j k}\right|}{X}}|1\rangle\right]
$, where $X=\frac{\Lambda_1}{\varepsilon d}$, and $X \geq \Lambda_{\max}\geq\|A\|_{\max }$
\State Apply $\mathcal{O}_A$ again, uncomputing the ancilla storing the value $A_{j k}$
\EndFunction
\end{algorithmic}
\end{algorithm}
This algorithm gives one query call to $\mathcal{O}_F$ and two query calls to $\mathcal{O}_A$.

Later work \cite{Low2019Qubitization} shows that $W$ can be understood as a block-encoding of $A$, as given in Definition \ref{def:block_encoding}.

\subsection{Block-encoding}
Block-encodings contain a matrix of interest $A$ within some known subspace of a unitary matrix. 
Herein, we construct block-encodings for $A$ which are suitable oracles for QLS-QSVT or our Hamiltonian simulation algorithm. 
Two popular choices for constructing these oracles are the linear combination of unitaries technique and the sparse access input model (SAIM). 
Recent work \cite{zhang2023circuit} showed that assuming that our input matrix is a linear combination of Pauli strings, we expect a linear combination of unitaries method to outperform SAIM \cite{zhang2023circuit}. 
However, SAIM uses the same set of oracles $\mathcal{O}_F, \mathcal{O}_A, $ as the functional QLS algorithms we consider. 
We will assume the QSVT oracle is prepared in this way so that QSVT and the functional algorithms remain comparable.

The general definition of a block encoding is given in Definition \ref{def:block_encoding}. When $A$ is sparse, then the block encoding of $A$ can be explicitly constructed \cite{Gilyen2019QuantumSingularValueTransformation}. We prefer to work with a measure encompassing both row ($d_r$) and column ($d_c$) sparsities, and define $d=\max\{d_c, d_r\}$, which then simplifies block-encodings for sparse matrices in  Lemma \ref{lemma:qsvt_BE_simple}. 
In our case, $A$ is assumed to be sparse and Hermitian so we will prepare a $(d, s+3, \varepsilon_{be})$ block encoding of $A$ with query complexity $3$ to $\mathcal{P}_A$.

\begin{definition}[Block encoding, Definition 43 in \cite{Gilyen2019QuantumSingularValueTransformation}]\label{def:block_encoding}
    \textit{Suppose that $A$ is an $s$-qubit operator. Then, given $\alpha, \varepsilon\in\mathbb{R}_+$ and $a\in\mathbb{N}$, we say that $(s+a)$-qubit unitary $U$ is a $(\alpha, a, \varepsilon)$ block encoding of $A$, if}
    \begin{equation}
        \left|\left|A_c-\alpha\left(\bra{0}^{\otimes a}\otimes I_s\right)U\left(\ket{0}^{\otimes a}\otimes I_s\right)\right|\right|\leq \varepsilon \, .
    \end{equation}
\end{definition}

\begin{lemma}[Block-encoding of sparse-access matrices, simplified from \cite{Gilyen2019QuantumSingularValueTransformation}]\label{lemma:qsvt_BE_simple}
    Let $A\in C^{2^s\otimes 2^s}$ be a matrix that is $d$-sparse, and $\left|\left|A\right|\right|_{\max}\leq 1$ . Denote the $b$-bit binary description of the $ij$th element of $A$ with $a_{ij}$.  $r_{ij}$ is the index of the $j$th non-zero entry of the $i$th row of $A$ (or if there are less than $i$ non-zero entries, then it is $j+2^s$.) $c_{ij}$ is the index of the $i$th non-zero entry of the $j$th column of $A$ (or if there are less than $j$ non-zero entries, then it is $i+2^s$).
    Suppose that we have access to the following sparse-access oracles
    \begin{align}
       \mathcal{O}_r:\ket{i}\ket{k}\rightarrow \ket{i}\ket{r_{ik}} \quad \forall i\in[2^s - 1], k\in[d] \, ,\\
        \mathcal{O}_c:\ket{l}\ket{j}\rightarrow \ket{c_{lj}}\ket{j} \quad \forall j\in[2^s - 1], l\in[d] \, ,\\
        \mathcal{O}_A:\ket{i}\ket{j}\ket{0}^{\otimes b}\rightarrow \ket{i}\ket{a_{ij}} \quad \forall i, j\in[2^s - 1] \, .
    \end{align}
    We can implement a $\left(d, s+3, \varepsilon\right)$ block encoding of $A$ with a single use of $\mathcal{O}_r, \mathcal{O}_c$ each, two uses of $\mathcal{O}_A$, and additionally using $\mathcal{O}\left(s+\log^{5/2}\left(\frac{d^2}{\varepsilon}\right)\right)$ while using $\mathcal{O}(b, \log^{5/2}(d^2/\varepsilon))$ ancilla qubits.
\end{lemma}

\section{\label{section:AppendixMI} QLS - QSVT Matrix Inversion}
\subsection{QSVT Basics}
Within QSVT, solving a system of linear equations is done by approximately implementing the matrix inverse.
Recall that every matrix has a singular value decomposition $A=V\Sigma W^{\dag}$, where $W, V$ are unitary and $\Sigma$ is a diagonal matrix with non-zero entries $\xi_1, \dots \xi_{r}$ where $r=\min(\rank(W), \rank(V))$.

Herein, assume $A$ is invertible and then we have
\[
A A^{-1}=V\Sigma W^{\dag}W{\Sigma}^{-1} V^{\dag}=I \, .
\]
From this we can see that the entries of $\Sigma^{-1}$, which are the singular values of $A^{-1}$, must be $\frac{1}{\xi_j}$. 
Applying a polynomial approximation of $1/x$ to the singular values of $A^\dag$ on a quantum device, we approximate the inverse of $A$\footnote{Note that In this case $A$ is Hermitian so the distinction between $A^{\dag}, A$ is immaterial. However, QSVT matrix inversion is often presented in the more general setting where $a\in\mathbb{M}^{n, m}$ where the algorithm prepares an approximation of the Moore-Penrose inverse. See \cite{Gilyen2019QuantumSingularValueTransformation} for details }. QSVT is precisely a framework for implementing polynomial transformation of singular values.
If the system is first initialized to $\ket{b}$, then the output will approximate the solution $\ket{x}$ of the linear system  with high probability.

We summarize QLS with QSVT in Alg. \ref{alg:qsvtinversion}. 
QSVT relies on access to the matrix $A^{\dag}$ in a block encoding, which in our case will be constructed with a few calls to sparse access oracles. 
Then a QSVT circuit (specifically the circuit in Lemma \ref{lemma:qsvtrealpoly}) approximates the target function discussed below. 
Our presentation is largely adapted from \cite{Gilyen2019QuantumSingularValueTransformation}, following \cite{lowthesis2017}. 
The target function is $\frac{1-\text{rect}(kx)}{2\kappa x}$, which can be made $\varepsilon$-close to the quantity we need on $\dom{k}$, for some $\varepsilon\in(0, 1)$. Moreover, it is bounded by $1$ on the interval $[-1, 1]$, and its polynomial approximation is given by the \textit{matrix inversion} (MI) polynomial $P_{\text{MI}}(x)$ in \autoref{sec:MI_poly_approx}.
We refer readers to \cite{Gilyen2019QuantumSingularValueTransformation} for a more detailed discussion.

\begin{algorithm}
\caption{QLS-QSVT from \cite{Gilyen2019QuantumSingularValueTransformation}}\label{alg:qsvtinversion}
    \begin{algorithmic}[1]
    \State Classical pre-processing: generate angles $\Phi(\kappa, \varepsilon)$ for $P_{\text{MI}}(x, \kappa, \varepsilon)$
        \Function{$\QSVT_{\text{MI}}$}{state $\ket{b}$, matrix $A$ with sparsity $d$, condition number $\kappa$, $\varepsilon >0$}
        \State Prepare $\ket{b}$ on state register with $\mathcal{P}_b$
        \State Run $\QSVT(\Phi(\kappa, \varepsilon), U_{A^{\dag}})$
        \State Run $\Amplify(\QSVT(\Phi, U_{A^{\dag}}), \cdot, \cdot)$
        \State Return $\ket{\psi_i}$, obeying $\left|\left|\frac{A^{-1}\ket{b}}{\left|\left|A^{-1}\ket{b}\right|\right|}-\ket{\psi_i}\right|\right|\leq \varepsilon$
        \EndFunction
    \end{algorithmic}
\end{algorithm}

\subsection{Pre-processing challenges}
We discuss QSP and QSVT pre-processing together here, because they require the same classical pre-processing step (QSP-processing) to determine a parameterized gate set from the input polynomial approximation. There are some techniques for doing this which are provably convergent with bounded error \cite{alexis2024infinitequantumsignalprocessing, Dong2022InfiniteQunatumSignalProcessing}.\\

In the case of the QLS problem, generating coefficient lists for the rectangle function can be problematic.  An alternative is to derive coefficients for a suitable approximation using, for example, the Remez algorithm \cite{dong_efficient_2021, Ying2022StableFacgorizationPhaseFactorsQSP}.  In fact, the largest published QSP-processing solutions use this to achieve QSP-angles for up to $\kappa=\mathcal{O}(10^2)$ and succeed to accuracy $\varepsilon=10^{-12}$. Application studies \cite{lapworth2024evaluationblockencodingsparse}  found that $\kappa=3000, \varepsilon=0.01$, is achievable, but that the cost of computing the Remez algorithm is still a significant challenge. For larger accuracy, $10^{-12}$,\cite{novikau_simulation_2023} found that only $\kappa=300$ was achievable; GPU parallelization of a Fourier series approximation of $1/x$ made solutions up to $\kappa=1000$ possible.\\ 

Our algorithm instead uses a closed form expression for an approximation polynomial, which are analytically bound to the target function. For this reason, our degree estimates are higher than may be feasible for given instances. To our knowledge, the largest successful angle-finding procedure with this analytically bounded-error approximation is \cite{Haah2019ProductDecompositionPeriodicFunctionsQSP}, which explicitly depends on arbitrary precision arithmetic. This approximation achieves the same asymptotic scaling, with $\kappa=1000$ instances up to an accuracy of $10^{-9}$. 

\subsection{QSVT for Real Polynomials}
The QSVT circuit is quite simple at a high level: we can implement a degree-n polynomial on the singular values of a matrix $A$, given $n$ steps and a block encoding $U_A$.
We have the following lemma for implementing a real polynomial with QSVT and a block encoding of $A$.
\begin{lemma}[QSVT by real polynomials, paraphrased from Corollary 18 in \cite{Gilyen2019QuantumSingularValueTransformation}]\label{lemma:qsvtrealpoly}
    Let $U_A$ be a block encoding of $A$ defined as in Lemma \ref{lemma:qsvt_BE_simple}, so that for projectors $\Pi,\Tilde{\Pi}$ and unitary $U_A$, $\Tilde{\Pi}U_A\Pi=A$.  Given an angle set $\Phi$, we define the \textit{alternating phase modulation circuit}, which is
\begin{equation}
\begin{split}
U_{\Phi}=e^{i\phi_1\left(2\Tilde{\Pi}-I\right)}U_A
\prod_{j=1}^{(n-1)/2}&\left(e^{i\phi_{2j}\left(2\Pi-I\right)}U_A^{\dag}\right.\\
&\cdot \left. e^{i\phi_{2j+1}\left(2\Tilde{\Pi}-I\right)}U_A\right)
\end{split}
\end{equation}
for $n$ odd and 
\begin{equation}
U_{\Phi}=\prod_{j=1}^{n/2}\left(e^{i\phi_{2j-1}\left(2\Pi-I\right)}U_A^{\dag}e^{i\phi_{2j}\left(2\Tilde{\Pi}-I\right)}U_A\right) 
\end{equation}
for $n$ even.
Now suppose that $P_{R}\in\mathbb{R}[x]$ is a degree $n$-polynomial satisfying
    \begin{itemize}
        \item $P_R$ has parity-$n \text{ mod }2$
        \item $|P_R(x)|\leq 1$  $\forall x\in[-1, 1]$
    \end{itemize}
    Then there exists $\Phi\in\mathbb{R}^n$ and a corresponding phase modulation circuit $U_{\Phi}$ building $P(x)\in\mathbb{C}$ such that $Re(P(x))=P_R(x)$. Using $\Phi$ and $\Phi_{-1}$ which builds $\left(P(x)\right)^*$, the operation applying $P_R$ to the singular values of $A$ is 
    \begin{align}
    P_R^{(SV)}\left(\Tilde{\Pi}U_A\Pi\right)    &=\left(\bra{+}\otimes\Tilde{\Pi}_{\text{parity}(n)}\right)\cdot \nonumber\\
    &\left(\ket{0}\bra{0}\otimes U_{\Phi}+\ket{1}\bra{1}U_{-\Phi}\right) \left(\ket{+}\otimes{\Pi}\right) \, ,
    \end{align}
    where 
    \begin{align}
        \Tilde{\Pi}_{0}&={\Pi} \, ,\\
        \Tilde{\Pi}_{1}&=\Tilde{\Pi} \, .
    \end{align}
\end{lemma}
Beginning from a real polynomial $P$ and obtaining suitable $\Phi$ is a complex pre-processing problem, often handled by optimization methods, see for example \cite{Dong2022InfiniteQunatumSignalProcessing, Ying2022StableFacgorizationPhaseFactorsQSP}.
The LCU step is extremely simple, and can be done with a single additional ancillary, two controlled-QSVT circuits, and two Hadamard gates. 

The query complexity of real-polynomial QSVT, counting $CU_A, U_A,$ and $U_A^{\dag}$ operations equally, is simply
\begin{equation}\label{eq:realqsvtcost}
    \cost{[U_{P_R}]}=n \, .
\end{equation}

In our case, $\Pi, \Tilde{\Pi}$ are generalized Toffoli gates over $3$ qubits (because the Toffoli gates are controlled on the block encoding ancillary qubit).
We do not distinguish between the cost of a controlled $1$ and a controlled $0$ gate. 
The QSVT procedure requires $\left\lceil\log_2 \dim(U_A)\right\rceil + 2$ ancillary qubits, where $U_A$ is the block encoding as in Lemma \ref{lemma:qsvt_BE_simple}. 

\subsection{Post-processing}
To demonstrate how post-processing affects the QSVT-QLS, we first consider QSP post-processing, which is easier, and then extend to the QSVT case. We are always using Hermitian matrices with eigenvalues $\lambda\in[-1, 1]$, and so we can always map the eigenvalues to $z=e^{i\arccos(\lambda)}$ on the unit circle, which is the variable QSP acts on. However, in this section, we will always write argument $\lambda$ for simplicity and consistency between the QSP and QSVT cases. Given an ancillary qubit $\ket{0}$ and unitary $U$ with eigenvectors $\ket{\lambda}$, QSP with complex functions performs the following operation on input state $\ket{0}\ket{\lambda}$, 
\begin{equation}
    \ket{0}\ket{\lambda}\rightarrow f(\lambda)\ket{0}\ket{\lambda}+g(\lambda)\ket{1}\ket{\lambda},
\end{equation}
and QSP with real functions performs
\begin{equation}
    \ket{+}\ket{\lambda}\rightarrow f(\lambda)\ket{+}\ket{\lambda}+g(\lambda)\ket{-}\ket{\lambda} .
\end{equation}
In both cases the construction guarantees that $\left|g(x)\right|^{2}+\left|f(x)\right|^{2}=1$ for all $x\in [-1, 1]$\footnote{For simplicity we sacrifice some notational precision by not distinguishing between the complex QSVT circuit and the real LCU of QSVT circuits. In either case, the important point is that the output QSVT state is normalized.}.\\

Consider now the action on a general initial state $\ket{b}=\sum_i\beta_i\ket{u_i}$ decomposed in the $U$ eigenbasis, where $\sum_i|\beta_i|^2=1$.
Then we have the straightforward generalization
\begin{align}
    \sum_i\beta_i\ket{+}\ket{u_i}&\rightarrow  \sum_i\beta_if(\lambda_i)\ket{+}\ket{u_i}\nonumber\\
    &+ \sum_i\beta_ig(\lambda_i)\ket{-}\ket{u_i}=\ket{\QSP(b)}.
\end{align}

By a straightforward calculation, the success probability of preparing the  normalized output $\frac{\sum_i\beta_if(\lambda_i)\ket{u_i}}{\norm{\sum_i\beta_if(\lambda_i)\ket{u_i}}}$ after post-selection on $\ket{+}$ is $\norm{\sum_i\beta_if(\lambda_i)\ket{u_i}}^2$.\\

We now generalize to the case where $U$ is a $(\alpha_{BE}, a, 0)$ block encoding of a Hermitian matrix prepared with Lemma \ref{lemma:qsvt_BE_simple}. Now, $U=\frac{1}{\alpha_{BE}}\ket{0^{\otimes a}}\bra{0^{\otimes a}}\otimes A + \Tilde{A}$ where $\Tilde{A}$ is a matrix with the same dimension of $U$ such that $U$ is unitary and $\bra{0^{\otimes a}}\tilde{A}\ket{0^{\otimes a}}=0$. To utilise this block encoding in real QSVT with length $n$, we assume an initial preparation of $\ket{+}\ket{0^{\otimes a}}\ket{0}$, and then apply the QSVT unitary. Now, QSVT prepares
\begin{align}
    \sum_i\beta_i\ket{+}\ket{0^{\otimes a}}\ket{u_i}\rightarrow &\sum_i\beta_if\left(\frac{\lambda_i}{\alpha_{BE}}\right)\ket{+}\ket{0^{\otimes a}}\ket{u_i}\nonumber\\
    + \, &\sum_i\beta_ig\left(\frac{\lambda_i}{\alpha_{BE}}\right)\ket{-}\ket{0^{\otimes a}}\ket{u_i}\nonumber\\
    + \, &U_{\Phi}\Tilde{A}\ket{+}\ket{0^{\otimes a}}\ket{b}=\ket{\QSVT(b)} .
\end{align}
A straightforward calculation verifies that $\norm{\ket{\QSVT(b)}}=1$. See \cite{martyn_grand_2021, dong_efficient_2021, Gilyen2019QuantumSingularValueTransformation} for examples of block encodings commonly used in QSVT.

The success probability of obtaining normalized output $\frac{\sum_i\beta_if\left(\frac{\lambda_i}{\alpha_{BE}}\right)\ket{u_i}}{\norm{\sum_i\beta_if\left(\lambda_i\right)}}$ after post-selection on $\ket{+}\ket{0^{\otimes a}}$ is $\norm{\sum_i\beta_if\left(\frac{\lambda_i}{\alpha_{BE}}\right)}^2$.\\

\subsection{Functional Approximation}\label{sec:MI_poly_approx}
Because $\frac{1}{1/x}>1$ on $[-1,1]$, we need a normalized target function. One can easily verify that the following function is proportional to $1/x$ on $\dom{\kappa}$ and bounded by one on $[-1, 1]$
\[
\frac{1-\rect(\kappa x/2)}{\kappa x} \, .
\]
We require analytically bounded functional approximations to the rectangle and inverse functions\footnote{We ignore possible Remez algorithm approaches to approximating the inverse function, see \cite{Dong_2021} for more details.} which fulfills the criteria from Lemma \ref{lemma:qsvtrealpoly}. 
Another difficulty is that the approximation to the rectangle function will be invalid on pieces of the unit circle, where it fluctuates within $[0, 1]$. 
So we need to ensure that our $1/x$ approximation is already $\leq1$ on these pieces - the easiest way to do so is to set the invalidity in $[\frac{-1}{\kappa}, \frac{-1}{2\kappa}]\cup[\frac{1}{2\kappa}, \frac{1}{\kappa}]$ and approximate
\[
\frac{1-\rect(\kappa x)}{2\kappa x} \, .
\]

Beginning from Eq. \eqref{eq:chebyshev_fcn}, we have precisely such an $\varepsilon$-close approximation to the inverse function on $\dom{k}$, $g(x)$. \cite{lowthesis2017} derives the following analytically bound approximation for $\rect(\frac{x}{2t})$:
\begin{align}
 p_{\text{rect}, t, \delta, \varepsilon }(x) &=\frac{1}{2}\left(p_{\text{sign},K, \varepsilon }(x+a)-p_{\text{sign},K, \varepsilon }(x-a)\right) \, ,
\end{align}
where $a=t+\frac{\delta}{4}$, and $K=\frac{\delta}{2}$.
$t$ is half the width of the interval, and $\delta$ is the width of the sections of the unit circle where the approximation is invalid, $[-t-\delta/2, -t]\cup [t, t+\delta/2]$.
This uses the following approximation of the sign function.
\begin{lemma}[Polynomial approximation to $\textrm{sign}$, paraphrased from 6.25-6.27 in \cite{lowthesis2017}]\label{lemma:sign_approx}
    Given $K>0, a\in [-1, 1], \varepsilon\in (0, 2\sqrt{2/e\pi})$, the sign function $\textrm{sign}(x-a)$ has an $\varepsilon$-close approximation on $x\in [-1, a-K/2]\cup [a+K/2, 1]$, given by $p_{\textrm{erf}}((x-a)/2, 2k, n)$
    \begin{align}
   {p}_{\text{sign}}\left(x, a,K,  \varepsilon\right)&=p_{\text{erf}}\left(\frac{x-a}{2}, 2k, n\right)\label{eq:low_sign_approx}\\
    n_{\text{rect}}&=2n_{\text{exp}}\left(2k^2, \frac{\sqrt{\pi}\varepsilon}{16k}\right)+1 \\
    k&=\frac{\sqrt{2}}{K}\log^{1/2}\left(\frac{8}{\pi\varepsilon^2}\right) \\
    \nonumber
    p_{\text{erf}}(x, k, n)&=\frac{2k e^{-\frac{k^2}{2}}}{\sqrt{\pi}}\left[(-1)^{\frac{n-1}{2}}I_{\frac{n-1}{2}}\left(\frac{k^2}{2}\right)\frac{T_{n}(x)}{n}\right.\\ 
    \nonumber
    &\left.+\sum_{j=0}^{(n-3)/2}(-1)^{j}\left(I_{j}\left(\frac{k^2}{2}\right)+I_{j+1}\left(\frac{k^2}{2}\right)\right)\right.\\
    &\left.\cdot\frac{T_{2j+1}(x)}{2j+1}\right]\\
    n_{\text{exp}}\left(\beta, \varepsilon\right) &= \left\lceil\sqrt{2\log(\frac{4}{\varepsilon})\left\lceil\max\left(\beta e^2, \log(\frac{2}{\varepsilon})\right)\right\rceil}\right\rceil.
\end{align}
\end{lemma}

\subsubsection{MI polynomial bounds}\label{sec:MI_function_bounds}
\begin{lemma}[Matrix Inversion Polynomial Approximation]\label{lemma:MI_function_bounds}
    We define a composite polynomial 
    \begin{equation}\label{eq:MI_poly_def}
   P_{MI, \varepsilon, \kappa}(x)= \frac{\left(1-{P_{\text{rect}, \frac{1}{2\kappa}, \frac{1}{\kappa}, \varepsilon_{\text{rect}}}(x)}\right)}{1+\varepsilon_{\text{inv}}}\frac{P_{1/x, \varepsilon_{\text{inv}}, \kappa}(x)}{2\kappa} \, ,
\end{equation}
where $P_{1/x, \varepsilon_{inv}, \kappa}(x), P_{\text{rect}}$ are defined respectively in Eq. \ref{eq:chebyshev_fcn} ($g(x)$) and Lemma \ref{lemma:sign_approx}. Define $\varepsilon_{\text{inv}}=\frac{\varepsilon}{2}$ and $\varepsilon_{\text{rect}}=\min\left(\frac{\varepsilon}{2},\frac{\kappa}{2j_0} \right)$. Then $P_{MI}(x)$ is $\varepsilon$-close to $\frac{rect(\kappa x)}{2\kappa x}$ on $\dom{\kappa}$.
\end{lemma}
\begin{proof}
    First, we require $P_{MI, \varepsilon, \kappa}$ to be less than $1$ on $x\in [-1, 1]$. Accounting for how bounds on the two composite functions vary over $[-1, 1]$, we have the following
\begin{align}
    \left|P_{MI, \varepsilon, \kappa}\right|&\leq\frac{1}{2\kappa(1+\varepsilon_{\text{inv}})}\left|P_{1/x, \varepsilon_{\text{inv}}, \kappa}(x)\right|\nonumber\\
    &\cdot \left|1-{P_{\text{rect}, \frac{1}{2\kappa}, \frac{1}{\kappa}, \varepsilon_{\text{rect}}}(x)}\right|\\
    &\leq \begin{cases}
          \frac{1}{2\kappa}|\kappa +\varepsilon_{\text{inv}}|\cdot |1-0 + \varepsilon_{\text{rect}}| \, , \nonumber\\
          \quad \quad|x|>\frac{1}{\kappa}\\
         \frac{1}{2\kappa(1+\varepsilon_{\text{inv}})}|\kappa +\varepsilon_{\text{inv}}|\cdot \left(1+\left|{P_{\text{rect}, \frac{1}{2\kappa}, \frac{1}{\kappa}, \varepsilon_{\text{rect}}}(x)}\right|\right), \nonumber\\
          \quad \quad \frac{1}{2\kappa}<|x|<\frac{1}{\kappa}\\
         \frac{1}{2\kappa(1+\varepsilon_{\text{inv}})}\left|P_{1/x, \varepsilon_{\text{inv}}, \kappa}(x)\right|{\varepsilon_{\text{rect}}} \, , \nonumber\\
          \quad \quad  |x|<\frac{1}{2\kappa}.
    \end{cases}
\end{align}
In the first case, $|x|>\frac{1}{\kappa}$,
\begin{align}
   \left|P_{MI, \varepsilon, \kappa}\right|&\leq  \frac{1}{2\kappa(1+\varepsilon_{\text{inv}})}|\kappa +\varepsilon_{\text{inv}}|\cdot |1-0 + \varepsilon_{\text{rect}}|\\
   &\leq  \frac{1}{2\kappa(1+\varepsilon_{\text{inv}})}(\kappa +\varepsilon_{\text{inv}})\cdot(1+\varepsilon_{\text{rect}})\\
   &\leq \frac{1}{2}(1 +\varepsilon_{\text{inv}})\\
   &\leq 1
\end{align}
whenever $\varepsilon_{\text{inv}}\geq \varepsilon_{\text{rect}}$, which becomes the first condition.
In the next range, $P_{1/x}$ is still a valid approximation but now $P_{\text{rect}}$ is not. However, from \cite{Gilyen2019QuantumSingularValueTransformation} we know that this function is bound by $1$ on the whole domain $[-1, 1]$, so
\begin{align}
    \left|P_{MI, \varepsilon, \kappa}\right|&\leq  \frac{1}{2\kappa(1+\varepsilon_{\text{inv}})}|\kappa +\varepsilon_{\text{inv}}|\nonumber\\
    &\cdot \left(1+\left|{P_{\text{rect}, \frac{1}{2\kappa}, \frac{1}{\kappa}, \varepsilon_{\text{rect}}}(x)}\right|\right)\\
    &\leq \frac{2}{2\left(1+\varepsilon_{\text{inv}}\right)}\left(1+\frac{\varepsilon_{\text{inv}}}{\kappa}\right)\\
    &\leq \frac{1}{\left(1+\varepsilon_{\text{inv}}\right)}\left(1+{\varepsilon_{\text{inv}}}\right)\leq 1 \, .
\end{align}
We have chosen to subnormalize $P_{\text{rect}}$ with $(1+\varepsilon_{\text{inv}})$ to ensure this works out. Finally, in the $|x|<\frac{1}{2\kappa}$ range,
\begin{align}
     \frac{1}{2\kappa(1+\varepsilon_{\text{inv}})}\left|P_{1/x, \varepsilon_{\text{inv}}, \kappa}(x)\right|{\varepsilon_{\text{rect}}}\nonumber\\
     \leq  \frac{1}{2\kappa}\left|P_{1/x, \varepsilon_{\text{inv}}, \kappa}(x)\right|{\varepsilon_{\text{rect}}}\\
     \leq \frac{4j_0}{2\kappa}\varepsilon_{\text{rect}}\leq 1 \, .
\end{align}
This leads to the constraint $\varepsilon_{\text{rect}}\leq \frac{\kappa}{2j_0}$.
We also need to demonstrate that the approximation is $\varepsilon$-close to $\frac{1}{2\kappa x}$ on $|x|\in [\frac{1}{\kappa}, 1]$; we can ignore the rest of the domain.
\begin{align}
  &\left|\frac{1-\rect({\kappa x})}{2\kappa x}-P_{MI, \varepsilon, \kappa}\right|\\
    &= \left|\frac{1-\rect({\kappa x})}{2\kappa x}+\right.\nonumber\\
    &\left.(1-\rect(\kappa x))\frac{P_{\text{inv}, \kappa,\varepsilon_{\text{inv}}}}{2\kappa(1+\varepsilon_{\text{inv}})}\right.\nonumber\\
    &\left.-(1-\rect(\kappa x))\frac{P_{\text{inv}, \kappa,\varepsilon_{\text{inv}}}}{2\kappa(1+\varepsilon_{\text{inv}})}\right.\nonumber\\
    &\left.-P_{MI, \varepsilon, \kappa}\right|\\  
    &\leq \left|{1-\rect({\kappa x})}\right|\cdot\left|\frac{1}{2\kappa x}-\frac{P_{\text{inv}, \kappa,\varepsilon_{\text{inv}}}}{2\kappa(1+\varepsilon_{\text{inv}})}\right|\nonumber\\
    &+\left|(1-\rect(\kappa x))-P_{\text{rect}, \varepsilon_{\text{rect}}, \frac{1}{2\kappa},\frac{1}{\kappa}}(x)\right|\nonumber\\
    &\cdot\frac{\left|P_{\text{inv}, \kappa,\varepsilon_{\text{inv}}}\right|}{2\kappa(1+\varepsilon_{\text{inv}})}
\end{align}
\begin{align}
    &\leq \frac{\left|\frac{1}{x}-P_{\text{inv}, \kappa, \varepsilon_{\text{inv}}}\right|+\left|\frac{\varepsilon_{\text{inv}}}{x}\right|}{2\kappa(1+\varepsilon_{\text{inv}})} + \frac{\varepsilon_{\text{rect}}(\kappa+\varepsilon_{\text{inv}})}{2\kappa(1+\varepsilon_{\text{inv}})}\\
    &\leq \frac{\varepsilon_{\text{inv}}+\kappa\varepsilon_{\text{inv}}}{2\kappa(1+\varepsilon_{\text{inv}})} +\frac{\varepsilon_{\text{rect}}(\kappa+\varepsilon_{\text{inv}})}{2\kappa(1+\varepsilon_{\text{inv}})}\\
    &\leq \frac{4\kappa\varepsilon_{\text{inv}}}{2\kappa(1+\varepsilon_{\text{inv}})}\\
    &\leq 2\varepsilon_{\text{inv}} \, .
\end{align}
So we can set $\varepsilon_{\text{inv}}=\frac{\varepsilon}{2}$ and $\varepsilon_{\text{rect}}=\min\left(\frac{\varepsilon}{2},\frac{\kappa}{2j_0} \right)$.
\end{proof}

\subsection{Proof of Lemma \ref{lemma:CostQSVTMI}}

\begin{proof}
We begin by using Lemma \ref{lemma:qsvt_BE_simple} to construct a block encoding of $A$, which will have subnormalization $\alpha_{BE}=\frac{1}{d}$. This is equivalent to the statement that we have prepared a block encoding of $\frac{A}{d}$. Our QSVT oracle uses $4$ calls to the sparse access oracles $\mathcal{O}_F, \mathcal{O}_A$. 

We use rescaled $\kappa'=d\kappa$, to ensure that the functional approximation is valid on the new domain $\dom{d\kappa}$. From \autoref{lemma:MI_function_bounds}, we can prepare an $\varepsilon'=\frac{\varepsilon}{6\kappa}$-close approximation to $\frac{1}{2d\kappa (x/d)}$\footnote{Subnormalization in the block encoding of $A$ means that we are computing a functional approximation of $\frac{d}{x}$, rather than $1/x$.}. Implementing $P_{MI,\varepsilon',\kappa'}$ in QSVT with a block encoding of $\frac{A}{d}$, and after post-processing, we obtain a state {$\sum_i\beta_iP_{MI,\varepsilon',\kappa'}(\frac{\lambda_i}{d}){\ket{u_i}}$ }\footnote{From this point on, we will simply denote $P_{MI,\varepsilon',\kappa'}$ by $P_{MI}$}.  With the re-scaled accuracy, we compute the parameters in Lemma \ref{lemma:MI_function_bounds}, obtaining a polynomial of final degree $n_{MI}=n_{\text{rect}}+n_{1/x}$. Then the circuit cost is given by the odd case of equation \eqref{eq:realqsvtcost}, as $n_{MI}$, leading to an overall  query complexity of $4n_{MI}$ to $\mathcal{O}_F, \mathcal{O}_A$.
The QSVT routine succeeds with probability $p=\norm{\sum_i{\beta_i}P_{MI}\left(\frac{\lambda_i}{d}\right)}^2$. Although there are versions of amplitude amplification designed specifically in QSVT, we use Lemma \ref{lemma:CostQAA} for a better comparison to other algorithms in this work.

The smallest magnitude $|P_{MI}\left(\frac{\lambda_i}{d}\right)|$ can have on $\dom{d\kappa}$ is $|\frac{1}{2\kappa\lambda_i}|-\varepsilon'$, so we have
\begin{align}
      p_{0}&\geq \norm{\sum{\beta_i}\left(\left|\frac{d}{2d\kappa\lambda}\right|-\varepsilon'\right)}^2\\
      &\geq \norm{\sum{\beta_i}\left(\frac{1}{2\kappa}-\varepsilon'\right)}^2\\
      &=\left(\frac{1-2\kappa\varepsilon'}{2\kappa }\right)^2\\
      &=\left(\frac{1-\frac{\varepsilon}{2}}{2\kappa }\right)^2 \, .
\end{align}

It remains to show the resulting state will be $\varepsilon$-close to $\ket{x}$. We are guaranteed that 
\begin{equation}
    \abs{\frac{1}{2\kappa'x}-P_{MI}({x})}\le {\varepsilon}' \, ,
\end{equation}
from which we see that
\begin{equation}
    \abs{\frac{1}{2\kappa\lambda_i}-P_{MI}\left(\frac{\lambda_i}{d}\right)}\leq {\varepsilon}' \, .
\label{eq:unnormalizedfunctionbound}
\end{equation}
Now consider the bound between the following unnormalized states,
\begin{align}
     \left|\left|\frac{1}{2\kappa}A^{-1}\ket{b}-\sum_i\beta_iP_{MI}\left(\frac{\lambda_i}{d}\right){\ket{u_i}}\right|\right|\\
     =\left|\left|\frac{1}{2\kappa}\sum_i\frac{\beta_i}{\lambda_i}\ket{u_i}-\sum_i\beta_iP_{MI}\left(\frac{\lambda_i}{d}\right){\ket{u_i}}\right|\right|\\
    =\left|\left|\sum_i \beta_i\bigl(\frac{1}{2\kappa\lambda_i}-{P_{MI}\left(\frac{\lambda_i}{d}\right)}\bigr){\ket{u_i}}\right|\right|\\
    =\sqrt{\sum_i \abs{\beta_i}^2\bigl\vert\frac{1}{2\kappa\lambda_i}-P_{MI}\left(\frac{\lambda_i}{d}\right)\bigr\vert^2}\\
    \leq {\varepsilon}' \, .\label{eq:unnomalized_qsvt_state_bound}
\end{align}

However, we now need to show that Eq. (\ref{eq:unnomalized_qsvt_state_bound}) implies 
\begin{align}
    \norm{\ket{x}-\ket{\tilde{x}}}&\leq \varepsilon \, ,
\end{align}
where
\begin{equation}
    \ket{\tilde{x}} = \frac{{2\kappa}\sum_i{\beta_i}P_{MI}\left(\frac{\lambda_i}{d}\right)\ket{u_i}}{2\kappa\norm{\sum_i{\beta_i}P_{MI}\left(\frac{\lambda_i}{d}\right)\ket{u_i}}}.
\end{equation}

Using standard expansions, and defining the unnormalized state $\ket{P}={2\kappa}\sum_i\beta_iP_{MI}\left(\frac{\lambda_i}{d}\right){\ket{u_i}}$,
\begin{align}
    \norm{\ket{x}-\ket{\tilde{x}}}&\leq \norm{\ket{P}}\abs{\frac{1}{\norm{A^{-1}\ket{b}}}-\frac{1}{\norm{\ket{P}}}}\nonumber\\
    &+\frac{1}{\norm{A^{-1}\ket{b}}}\norm{A^{-1}\ket{b}-\ket{P}}\\
    &\leq \frac{1}{\norm{A^{-1}\ket{b}}}\left(\abs{{\norm{A^{-1}\ket{b}}}-{\norm{\ket{P}}}}\right.\nonumber\\
    &\left.+\norm{\ket{P}-A^{-1}\ket{b}}\right)\\
    &\leq\left(\abs{{\norm{A^{-1}\ket{b}}}-{\norm{\ket{P}}}}+\norm{\ket{P}-A^{-1}\ket{b}}\right) .\label{eq:qsvt-state-norm-bound}
\end{align}
From the inverse triangular inequality, we have
\begin{equation}
    \abs{\norm{A^{-1}\ket{b}}-\norm{\ket{P}}}
    \le \norm{A^{-1}\ket{b}-\ket{P}}\le \frac{\varepsilon}{3} \, .
\end{equation}

Substituting into Eq. \eqref{eq:qsvt-state-norm-bound}, and using Eq. \eqref{eq:unnomalized_qsvt_state_bound} to bound the state norm,
\begin{align}
    \norm{\ket{x}-\ket{\tilde{x}}}&\leq 2\norm{A^{-1}\ket{b}-\ket{P}}\\
    &\leq 4\kappa'\varepsilon'\leq \varepsilon \, .
\end{align}
Finally, the probability of success is
\begin{equation}
     p =\norm{\sum_i{\beta_i}P_{MI}\left(\frac{\lambda_i}{d}\right)}^2 \, ,
\end{equation}
which from Eq. \eqref{eq:unnormalizedfunctionbound} we can see is, up to order $\varepsilon$, 
\begin{equation}
    \frac{\norm{x}^2}{4\kappa^2} \, ,
\end{equation}
which is used in our numeric work for convenience.

\end{proof}

\section{\label{sec:LCU}Linear Combination of Unitaries}
Linear combination of unitaries (LCU) is an algorithm originally from \cite{Berry2015SimulationHamiltonianDynamicsTruncatedTaylorSeries, Berry2015HamiltonianSimulationNearlyOptimalDependenceOnAllParameters} which implements a linear combination of unitary operations on a quantum circuit. 
Alg. \ref{alg:lcu} is a version of LCU from \cite{Childs2017QuantumAlgorithmSistemyLinearEquations}, specialized to the quantum linear system problem. We state the result with a fixed success probability. 
\begin{algorithm}
\caption{LCU for QLS}\label{alg:lcu}
\begin{algorithmic}[1]
\Function{LCU}{device prepared in state $\ket{b}$, set of coefficients $\{\alpha_i\}$,  corresponding set of unitaries $\{U_i\}$, $\varepsilon >0$, $\Delta$ }
\State $U\gets\sum_{i=0}^{\Delta}\ketbra{i}\otimes U_i$, 
\State Apply $V$ to register $\ket{0^m}$
\Comment $V\ket{0}=\frac{1}{\sqrt{\alpha}}\sum_{i}\sqrt{\alpha_i}\ket{i}$, $\alpha=\sum_i\alpha_i$
\State Apply $U$
\State Apply $V^{\dag}$
\EndFunction
\end{algorithmic}
\end{algorithm}
\begin{lemma}[Linear Combination of Unitaries from \cite{Childs2017QuantumAlgorithmSistemyLinearEquations} ]
\label{lemma:lcu}
Let $A$ be a Hermitian operator with eigenvalues in a domain $\mathcal{D}\subseteq\mathbb{R}$. 
Suppose $f: \mathcal{D}\rightarrow \mathbb{R}$ satisfies $\norm{f(x)}\geq1$ for all $x\in\mathcal{D}$  and is $\varepsilon$-close to $\sum_i\alpha_iT_i$ indexed $i=0, \dots, \Delta$ on $\mathcal{D}$ for some $\varepsilon\in(0, 1/2)$, 
$\alpha_i>0$, and $T_i:\mathcal{D}\rightarrow\mathbb{C}$. 
Let $\{U_i\}$ be a set of unitaries such that 
$$U_i\ket{0^t}\ket{\phi}=\ket{0^t}T_i(A)\ket{\phi}+\ket{\Psi_i^{\perp}}$$
for all states $\ket{\phi}$ where $t\in\mathbb{Z}_+$ and $\ket{0^t}\bra{0^t}\otimes \mathcal{I}\ket{\Psi_i^{\perp}}=0.$ 
Given an algorithm $\mathcal{P}_b$ for preparing state $\ket{b}$, 
there is a quantum algorithm that prepares a quantum state $4\varepsilon$-close to $f(A)\ket{b}/\left|\left|f(A)\ket{b}\right|\right|$ and outputs a bit indicating whether it was successful or not. The success probability $\frac{\left|\left|f(A)\ket{b}\right|\right|^{2}}{\alpha^2}$ is lower bounded by $\frac{1}{\alpha^2}$. 
The number of queries to $O_F, O_A$ which prepare $A$ in a sparse-access oracle model is lower bounded by
\begin{equation}
    \cost[\lcu(\{U_i\}, \{\alpha_i\})]=\cost[U] \, ,
    \label{eq:LCU_cost}
\end{equation}
where
\begin{equation}
    U=\sum_{i=0}^{\Delta}\ketbra{i}\otimes U_i\,.
\end{equation}
\end{lemma}

\begin{proof}
For given coefficients $\alpha_i>0$, unitary operations $U_i$, and indices $i=0,\dots, \Delta$, define
\begin{equation}
    M=\sum_{i=0}^{\Delta}\alpha_i U_i\,,
\end{equation}
where $M$ is $\varepsilon$-close to $f(A)$ by assumption.
In general $M$ is not a unitary, hence we can apply this transformation to the state $\ket{0^m}\ket{b}$ probabilistically. 
To this end, we define a unitary $V$ acting on an ancillary register such that
\begin{equation}
V\ket{0}=\frac{1}{\sqrt{\alpha}} \sum_{i=0}^{\Delta}\sqrt{\alpha_i}\ket{i} ,
\end{equation}
where $\alpha=\sum_i\alpha_i$, and the controlled unitary 
\begin{equation}
    U=\sum_{i=0}^{\Delta}\ketbra{i}\otimes U_i \, .
\end{equation}
One can verify that
\begin{equation}
    V^{\dagger}UV\cdot (\ket{0^t}\otimes \ket{0^m}\ket{b})=\frac{1}{\alpha}\ket{0}\otimes M\ket{0^m}\ket{b}+\ket{\Psi^{\perp}},
\end{equation}
where  $(\ketbra{0}\otimes\mathds{1})\ket{\Psi^\perp}=0$. 
Therefore, the algorithm produces the desired state $f(A)\ket{b}/\left|\left|f(A)\ket{b}\right|\right|$ with probability $\frac{\left|\left|f(A)\ket{b}\right|\right|^{2}}{\alpha^2}$. From the condition $\left|\left|f(x)\right|\right|\geq 1$, the success probability has lower bound $\frac{1}{\alpha^2}$.\\

$\alpha$ and $\cost[U(\cost[U_i])]$ should have bounds specific to the given $f$. 
$V$ is implemented with a state preparation map \cite{Childs2017QuantumAlgorithmSistemyLinearEquations} on $m= \lceil\log (\Delta+1)\rceil$ qubits (note that the indexing starts at $0$) and \cite{Shende_2006} gives a procedure for computing this with $(\Delta+1)-2$ 2-qubit gates. Because these gates are all controlled rotations, further decompositions into CNOT and 1-qubit rotation gates would only increase the gate complexity by a linear factor \cite{nielsen2002quantum}, leading to
\begin{equation}
  \cost[V]\leq \left(\Delta-1\right)\left(4\cost_1+2\cost_2\right).
  \label{eq:costV}
\end{equation}
We have not argued this is an optimal state preparation scheme for $V$, but instead used a reasonable heuristic to justify why $\cost[V]$ is not expected to be the dominant contribution to the cost. Hence, it is dropped to obtain a lower bound and altogether the cost of LCU is given by 
\begin{equation}
    \cost[\lcu(\{U_i\},\{\alpha_i\})]=\cost[U] \, .
\end{equation}
\end{proof}

\end{document}